\documentclass{jfm}
\usepackage{graphicx}
\usepackage{epstopdf, epsfig}

\usepackage{subcaption}
\usepackage{nicefrac}
\usepackage{comment}
\usepackage{chemformula}
\usepackage{soul}

\usepackage{calc}
\usepackage{amsmath}
\usepackage{xcolor}

\newlength\dlf

\shorttitle{Shock cellular flame interaction}
\shortauthor{H. Yang and M. I. Radulescu}

\title{Dynamics of cellular flame deformation after a head-on interaction with a shock wave: reactive Richtmyer-Meshkov instability}

\author{Hongxia Yang\aff{1,2}
  \corresp{\email{yang.hongxia@foxmail.com}},
 \and Matei Ioan Radulescu\aff{2}}

\affiliation{\aff{1}Fire \& Explosion Protection Laboratory, Northeastern University, Shenyang, 110819, China
\aff{2}Department of Mechanical Engineering, University of Ottawa, 161 Louis-Pasteur, Ottawa, K1N 6N5, Canada}

\begin{document}

\maketitle

\begin{abstract}
Shock flame interactions are fundamental problems in many combustion applications ranging from flame acceleration to flame control in supersonic propulsion applications. The present paper seeks to quantify the rate of deformation of the flame surface and burning velocity caused by the interaction and to clarify the underlying mechanisms. The interaction of a single shock wave with a cellular flame in a Hele-Shaw shock tube configuration was studied experimentally, numerically, and theoretically. A mixture of stoichiometric hydrogen-air at sub-atmospheric pressure was chosen such that large cells can be isolated and their deformation studied with precision subsequent to the interaction. Following passage of the incident shock  and vorticity deposition along the flame surface, the flame cusps are flattened and reversed backwards into the burnt gas. The reversed flame then goes through four stages. At times significantly less than the characteristic flame burning time, the flame front deforms as an inert interface due to the Ricthmyer-Meshkov instability with non-linear effects becoming noticeable. At times comparable to the laminar flame time, dilatation due to chemical energy release amplifies the growth rate of Ricthmyer-Meshkov instability. This stage is abruptly terminated by the transverse burnout of the resulting flame funnels, followed by a longer front re-adjustment to a new cellular flame evolving on the cellular time scale of the flame. The proposed flame evolution model permits to predict the evolution of the flame geometry and burning rate for arbitrary shock strength below the shock-induced auto-ignition point and flames with unit Lewis number in two-dimensions.
\end{abstract}


\section{Introduction}
\label{intro}

The interaction of shocks with flames is a fundamental problem of reactive compressible flows with applications to flame acceleration and flame control in propulsion systems. Experimental evidence suggests that the passage of a shock over the flame would first result in important deformation of the flame surface and subsequent enhancement of the burning rate \citep{markstein1957shock, rudinger1958shock, urtiew1968transverse, scarinci1993amplification, ciccarelli2010role, rakotoarison2019mechanism, wei2017effects}. Also, the shock reflected from walls or other reflective surfaces may result in multiple interactions, which would lead to more considerable flame acceleration and possibly the deflagration-to-detonation transition \citep{thomas2001experimental}. The generally very rapid interactions make it hard to quantify the enhancement of the burning rate, which depends intimately on the flame front surface and the thermodynamic properties of the flow field. 

Efforts have focused on laminar flames, due to the well-posedness of initial conditions.  An example is the classic work of \citet{markstein1957shock}, which investigated the head-on collision between an initially curved laminar flame and a planar shock. He attributed the mechanism for the flame reversion to be a modified version of Taylor instability. More recent studies also suggest the Richtmyer-Meshkov (RM) instability, which is generally referred to the impulsive counterpart of Rayleigh-Taylor instability, to be the main factor associated to the increase of the total burning velocity by the extension of the flame front after the shock-flame interaction \citep{ju1998vorticity, khokhlov1999interaction, oran2007origins, dong2008numerical, zhu2013three, jiang2016parameterization, massa2012linear}. 

The linear and non-linear growth of the \textit{inert} RM instability is now relatively well understood.  A quick review is given here, since the reactive growth explored in the present paper builds upon this. The RM instability is known to amplify the perturbation on the interface through the impulsive deposition of vorticity on the interface by the shock by the baroclinic torque mechanism, which results from the misalignment of the pressure gradient of the shock and the density gradient across the interface. Since the pioneering work of \citet{richtmyer1960taylor}, extensive experiments, simulations and theoretical models have been formulated to elucidate the interface deformation subject to the inert RM instability.  Reviews of these studies can be found in \citet{brouillette2002richtmyer} and \citet{zhou2017rayleigh2,zhou2017rayleigh1}. The basic configuration of a shock passing a sinusoidal interface is generally adopted in these studies. The amplitude growth is typically recognized as a key indicator to represent the interface deformation. For the case of a shock passing a sinusoidal interface separating a light and a heavy fluid, the initial linear growth of the interface amplitude is well predicted by the impulsive model of \citet{richtmyer1960taylor} for sufficiently small perturbation amplitudes. The growth rate of the interface amplitude $\eta^{\prime}$ is given by the impulsive model and takes the form $\eta^{\prime} = k A_{t} [u] \eta_0$, where $k$ is the wavenumber, $[u]$ is the velocity jump associated to the passage of the shock, $\eta_0$ is the initial amplitude and $A_t = \nicefrac{\rho_2 - \rho_1}{\rho_2 + \rho_1}$ is the Atwood number with $\rho_1$ and $\rho_2$ refer to the light and heavy fluid density. Conversely, when the shock wave is directed from the heavy fluid to the light one, the interface amplitude decreases before reversing its phase and growing linearly. This linear growth relation has been proposed by \citet{meyer1972numerical} with an improvement to the impulsive model (MB model) by using the average of the amplitudes before and after the interaction along with the post-shock Atwood number. With the amplification of the amplitude progresses to be comparable to the wavelength, non-linear effects become significant. The non-linear mechanism leads to a decrease in the growth rate. Numerous models \citep{sadot1998study, mikaelian2003explicit, zhang1999quantitative, kilchyk2013scaling, zhang2016universality} have been developed to predict the amplitude growth in the nonlinear stage for arbitrary Atwood numbers, the late time non-linear growth rate of the interface was found to decay at the rate of $O(1/t)$. \citet{mikaelian2003explicit} proposed a model to join the linear and non-linear regime by making an abrupt change in the growth rate for arbitrary Atwood numbers. The model was found to show good agreement to other studies \citep{goncharov1999theory,oron2001dimensionality}.

Studies on the \textit{reactive} RM instability have demonstrated that the flame deformation induced by the single shock-laminar-flame interaction can be assumed to be in the laminar flamelet regime \citep{ju1998vorticity, khokhlov1999interaction,lutoschkin2013pressure}. The main features and the time scale of the initial flame evolution are similar to that of the inert case. While profound differences were noted in the later non-linear stages, the perturbation growth of the interface was found to be larger than that of the inert interface instability case due to both the interface stretching and flame propagation mechanism. As a result, the non-linear model for the inert RM instability would fail to predict the interface evolution for the reactive cases. Furthermore, the competition between the initial torque by the shock and the chemistry-induced baroclinic effect leads to the disappearance of small scale perturbations in the reaction cases than in the inert shock interface interaction cases \citep{khokhlov1999interaction, massa2012linear,wijeyakulasuriya2014analyzing}. \citet{scarinci1993amplification} evaluated the burning rate increase for corrugated flames with not too large initial amplitude through a series of experiments by passing the shock wave across a pre-defined corrugated flame front. By the inspection of the pressure records and conservation laws arguments, the different mechanisms of the thermodynamic dependence of flame speed, the increase of transport rates by interaction generated turbulence and deformation of the flame surface that might have influenced the resultant burning rate increase was examined. In the analysis, they attributed the dominant mechanism to the RM instability through the increase of the burning interface and the turbulent energy acting on the flame interface. The effect of chemistry was found to be more important at the later stage of the interaction, especially for higher Mach numbers \citep{chen2018numerical}. The results from the quasi-one-dimensional investigation of the problem by separating the one-dimensional effects of gas compression and two-dimensional flame front distortion also found the gas compression to be the dominant effect in influencing the total burning rate increase for stronger incident shocks \citep{scarinci1993amplification, kilchyk2013scaling, lutoschkin2013pressure}. However, the mechanism on the competition between the aforementioned potential effects at different stages is still not well understood. 

While previous work on shock-flame interactions reviewed above treated either a laminar flame with small perturbations for well-posedness, or the turbulent or strained flames constrained by the tube geometry for practical considerations as in the work of \citet{markstein1957shock} and  \citet{thomas2001experimental}, the present study addresses cellular flames.  Indeed, it is well known that planar laminar flames are unstable to both hydrodynamic (Landau-Darrieus) and thermo-diffusive instabilities \citep{sivashinsky1983instabilities}, which make them saturate to a cellular structure of finite amplitude \citep{sharpe2006nonlinear}.  Flames in practice where shock-flame interactions may be of importance, as in astrophysical, engine or large-scale accidental explosion scenarios are likely cellular due to the much larger characteristic propagation distance compared with the laminar flame thickness.   These cellular shapes are taken also conveniently as the initial seeds of the RM-type instability induced by the passage of the planar shock.  The objective of this paper is thus to quantitatively clarify the controlling mechanism on the flame front deformation for the fundamental problem of head-on interaction of a shock with a cellular unstable flame experimentally, numerically and theoretically.  The experiments are performed at low pressure in stoichiometric hydrogen-air, such that large cells can be isolated and their deformation studied with precision after passage of a shock wave. Also, the interaction can be readily amenable to numerical simulation and theoretical modeling. Similar to the inert RM instability studies, in the experiments, we take the flame front amplitude as an indicator for the evolution of the flame surface, under the impulsive action of the shock wave leading to RM instability. Numerical simulations are conducted to extend the observational time limited in the experiments. A simplified model is then proposed to predict the evolution of the flame geometry and burning rate for arbitrary shock strength below the shock-induced auto-ignition point and flames with unit Lewis number in two-dimensions.

The remainder of the present paper is organized as follows. Section \ref{sec:experimental procedure} provides a brief description of the experimental details. The experimental results of the stoichiometric hydrogen-air flame and the head-on interactions of the cellular flame with various incident shocks are given in section \ref{sec:experimental results}. Sections \ref{sec:numerical method and physical model} and \ref{sec:numerical results} present the numerical method and physical model to reconstruct the observed experimental flow field. Subsequent discussion on the evolution of the flame deformation following the interaction with the incident shock is presented in section \ref{sec:discussions}. Finally, conclusions are given in section \ref{sec:conclusion}.

\section{Experimental procedure}
\label{sec:experimental procedure}

\begin{figure}
	\centering
	\includegraphics[width=1\columnwidth]{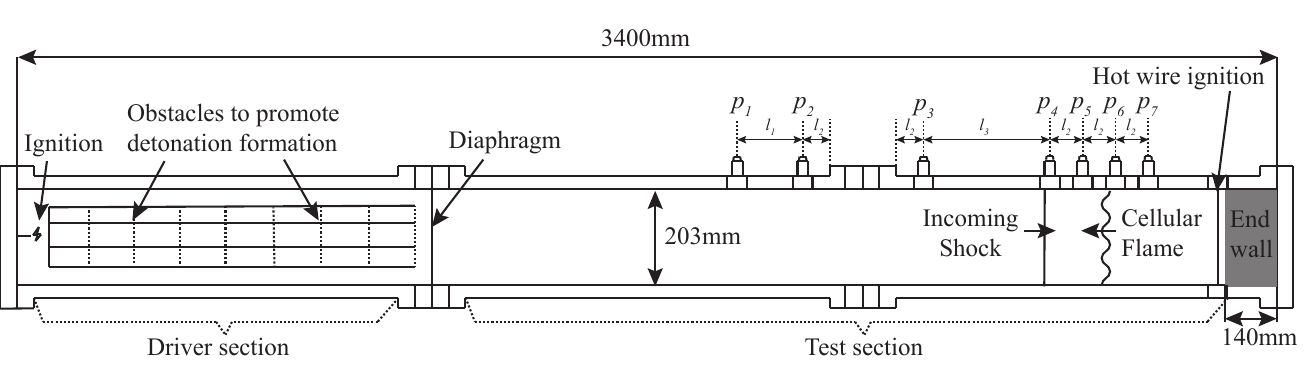}
	\caption{Schematic of the experiment setup in the shock tube, $l_{1}=203$ mm, $l_{2}=102$ mm, $l_{3} = 406$ mm. }
	\label{shock_tube}
\end{figure}

The experiments were conducted in a shock tube with dimensions of 3400 mm $\times$ 19.1 mm $\times$ 203.2 mm, as shown in figure\ \ref{shock_tube}. The flame was ignited by a 0.15-mm-thick hot tungsten wire mounted at the right end of the shock tube, which consisted of the test mixture of stoichiometric hydrogen-air. A reactive driver gas (\ch{C2H4}/3\ch{O2}) generated a shock wave propagating to the right, while the flame that was ignited at the opposite end has propagated a sufficient distance to acquire the desired cellular structure in the test mixture. The driver gas was separated from the test gas by a layer of aluminum sheet serving as diaphragm, which isolated the gases from each other prior to an experiment and avoided contamination of the gases.  The shock was initiated by triggering a detonation in the driver gas. This incident detonation broke the aluminum diaphragm and transmitted as a shock wave into the test gas, followed by a much weaker flame that did not participate in the experiment. The transmitted shock then traveled toward the test flame and interacted head-on. The strength of the incident shock was controlled by varying the initial pressure of the driver gas.

A series of high-frequency piezoelectric PCB pressure sensors (p$_1$-p$_7$) were mounted flush on the top wall of the shock tube to collect pressure signals and the arrival of the shock. The pressure signals were sampled at a rate of 1.538 MHz, and low-pass filtered at 100 kHz. The incident shock Mach number was evaluated using the pressure amplitude measured at each gauge and confirmed by the time of arrival. In the experiments, the incident shock Mach number was within a range of 1.5 to 1.9, which is much smaller than the critical shock Mach number of 3.7 that corresponds to an ignition delay of 1ms, deemed the auto-ignition limit in our facility. Also, the highest shock Mach number was chosen to avoid exceeding the maximum allowable operating pressure of the facility for safety concerns. For each experiment, the shock tube was evacuated to a pressure less than 80 Pa. Then the \ch{C2H4}/3\ch{O2} mixture was filled into the driver section before filling the rest part with 2\ch{H2}/air mixture from the opposite end of the sock tube. The gases were prepared in mixing tanks by the method of partial pressure and left to mix for more than 24 hours. The initial pressure of the test mixtures was in a range of 17 kPa to 21 kPa. The ambient temperature was 294K. A pair of optical quality glass windows was installed at the test section in order to visualize the phenomenon.  A Z-type Schlieren system with a field of view of 317.5 mm was implemented to capture the shock-flame interaction and the ensuing evolution. The image sequence was recorded using a high-speed camera (Phantom v1210). The frame rate was 59,590 frames per second (fps) with a resolution of 512 $\times$ 320, and the exposure time was set to 0.468 $\mu$s. 

\section{Experimental results}
\label{sec:experimental results}

\subsection{Flame evolution before the interaction}
\begin{figure}
	\centering 
	\includegraphics[width=0.7\linewidth]{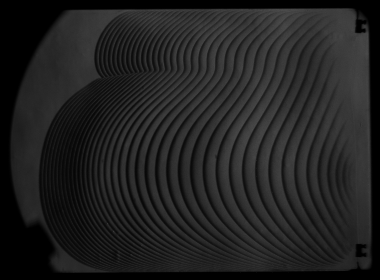}
	\caption{Superposition of flame fronts development at different time for stoichiometric hydrogen air ignited by hot tungsten wire. Pressure of 20.7 kPa. Recorded at 59,590 fps. The time interval between the flame front is 0.25 ms.}
	\label{flame}
\end{figure}

\begin{figure}
	\centering
	\subcaptionbox{\label{flame_later_frame} }{\includegraphics[width=0.72\linewidth]{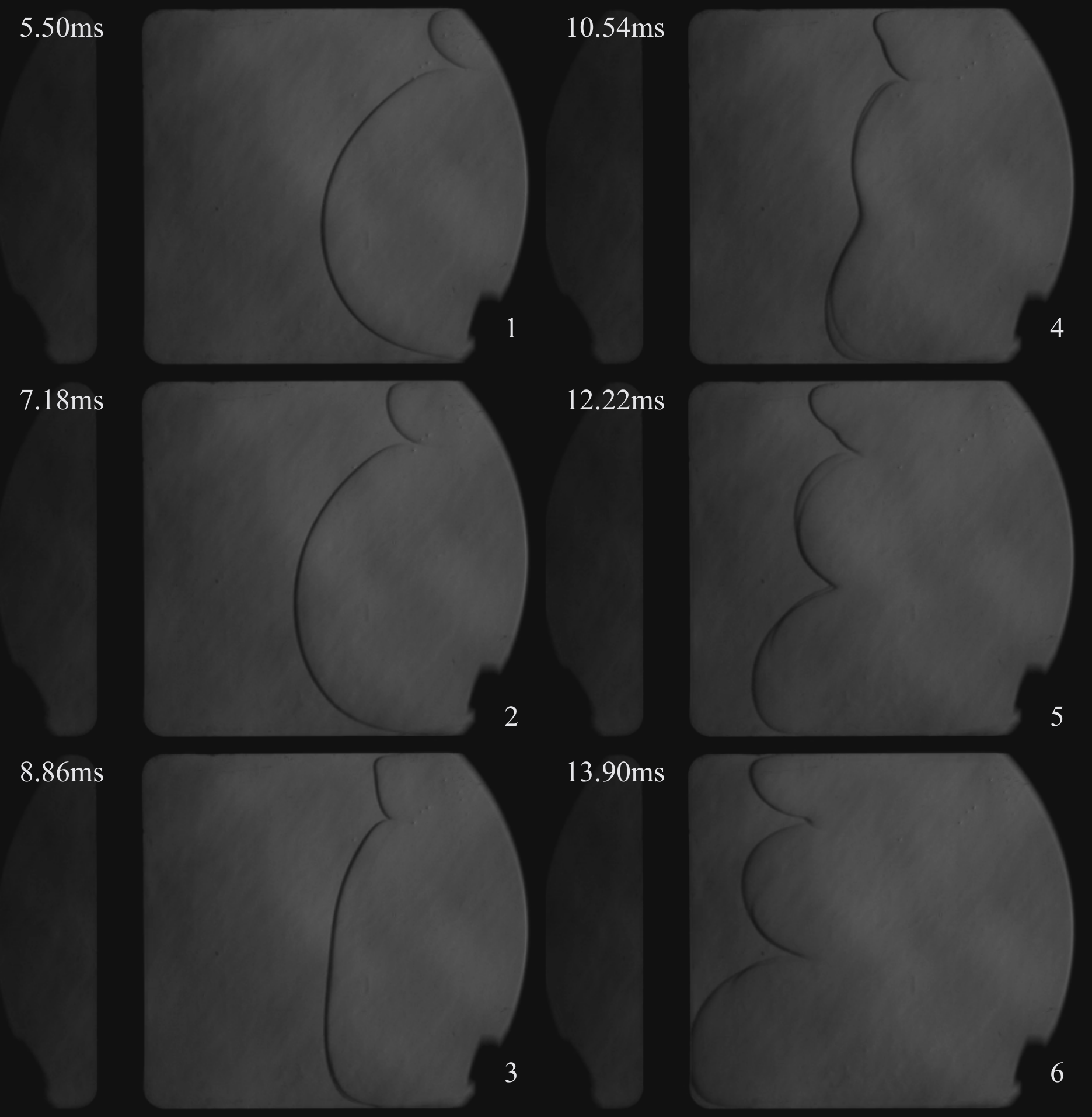}}
	\subcaptionbox{\label{flame_later_xt} }{\includegraphics[width=0.72\linewidth]{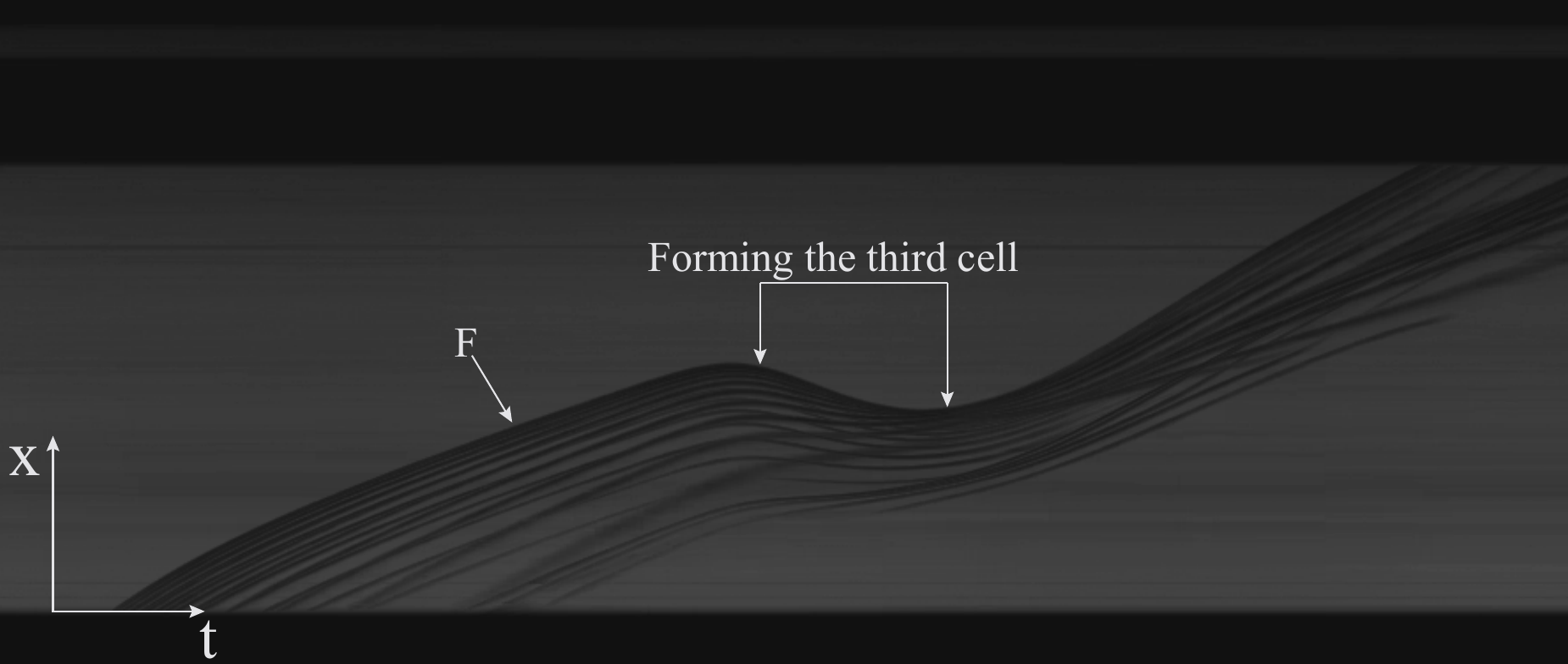}}
	\caption{(a) Schlieren image sequence, (b) space-time diagram obtained every 10th pixels for stoichiometric hydrogen-air cellular flame development. Pressure of 17.2 kPa. Recorded at 59,590 fps.}
	\label{flame_later}
\end{figure}

Figure \ref{flame} shows an example of the cellular flame establishment prior to its interaction with the shock. The superimposed Schlieren image illustrates the stoichiometric $\text{H}_{\text{2}}/\text{air}$ flame fronts evolution for the very early stage at an initial pressure of 20.7 kPa with a time interval of 0.25 ms. The flame propagated from right towards left. It can be seen from the variation of the distance between the flame fronts acquired in sequential frames in figure \ref{flame} that, once ignited by the vertical hot tungsten wire in the right end, the flame first slowly accelerated as the perturbation curvature increased, and then decelerated to form two cells. As the newly formed cells propagated forward, the increase of their amplitudes is noticeable. Figure\ \ref{flame_later} further demonstrates the birth of new cells from the bottom large cell ensuing the formation of two cells, with an initial pressure of 17.2 kPa. The flame front evolution was tracked every 10th pixels horizontally in each of the sequential frames, as shown in the space-time diagram in figure \ref{flame_later_xt}. Evident from the nearly linear slope of the flame front streaks in figure \ref{flame_later_xt} is that the flame propagated with practically steady speed before the formation of the third cell, and it showed the same trend after the fully development of three cells. The interactions of shock and three-cells-flame with the initial pressure of 17.2 kPa are reported as follows.

\subsection{Interaction of $M_\text{s}$ = 1.9 shock and the flame}
\begin{figure}
	\centering
	\includegraphics[width=0.8\linewidth]{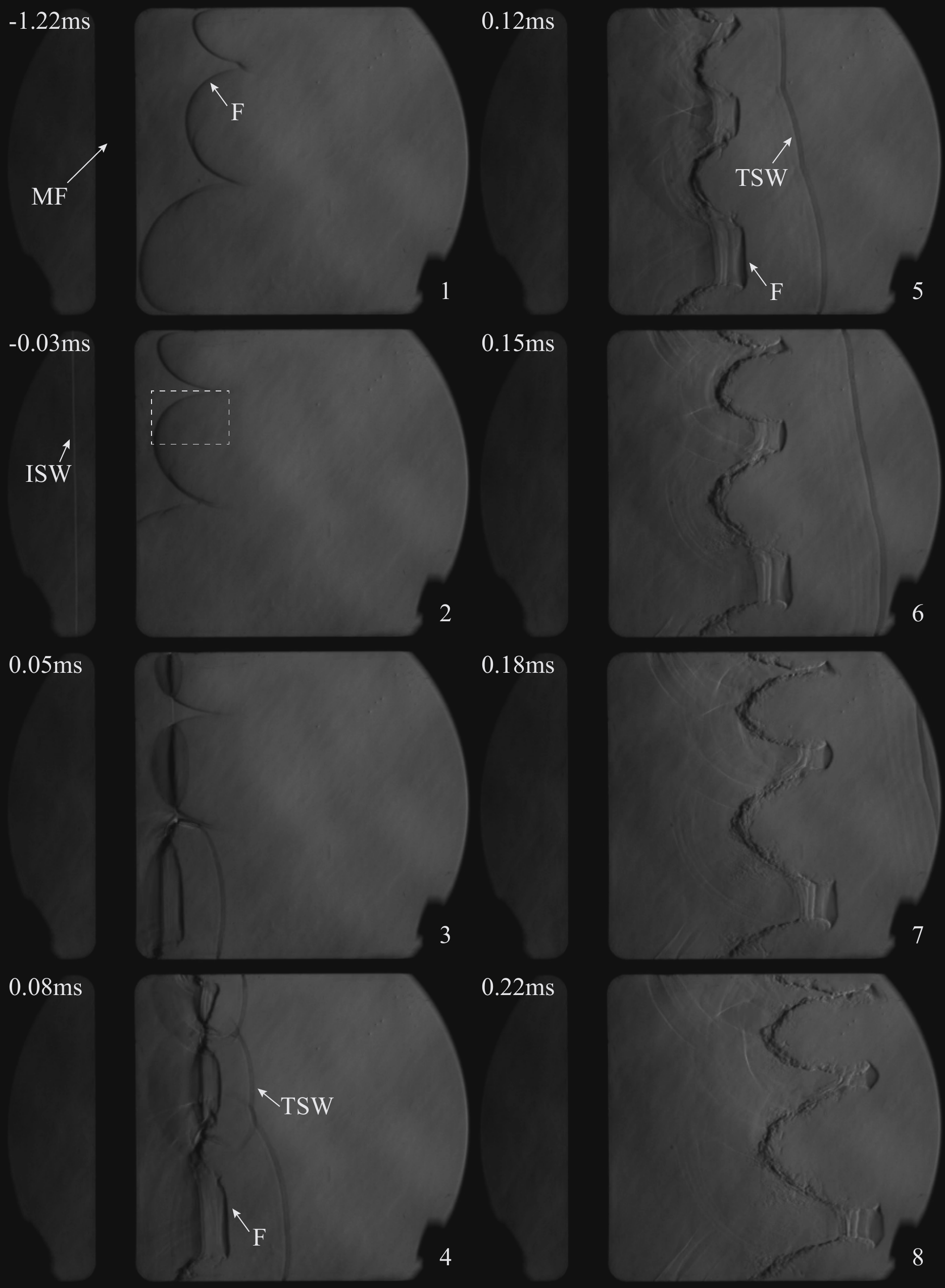}
	\caption{Schlieren image sequence of the interaction of a $M_\text{s}$ = 1.9 incident shock wave(ISW) with stoichiometric hydrogen-air flame(F). TSW is the transmitted shock wave, MF is the middle flange of the shock tube. Time 0 corresponds to the beginning of the interaction.}
	\label{M1_9interaction}
\end{figure}

\begin{figure}
	\centering
	\includegraphics[clip,width=0.6\columnwidth]{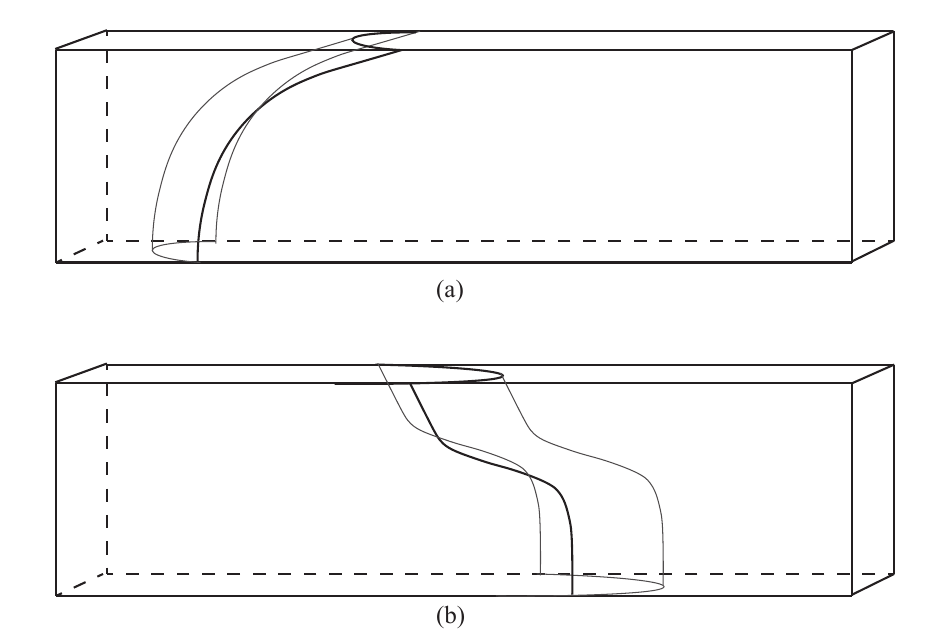}
	\caption{Sketch of the three-dimensional flame configuration (a) before and (b) after the interaction with the shock.}
	\label{3d_flame_sketch}
\end{figure}

\begin{figure}
	\centering
	\includegraphics[clip,width=0.4	\columnwidth]{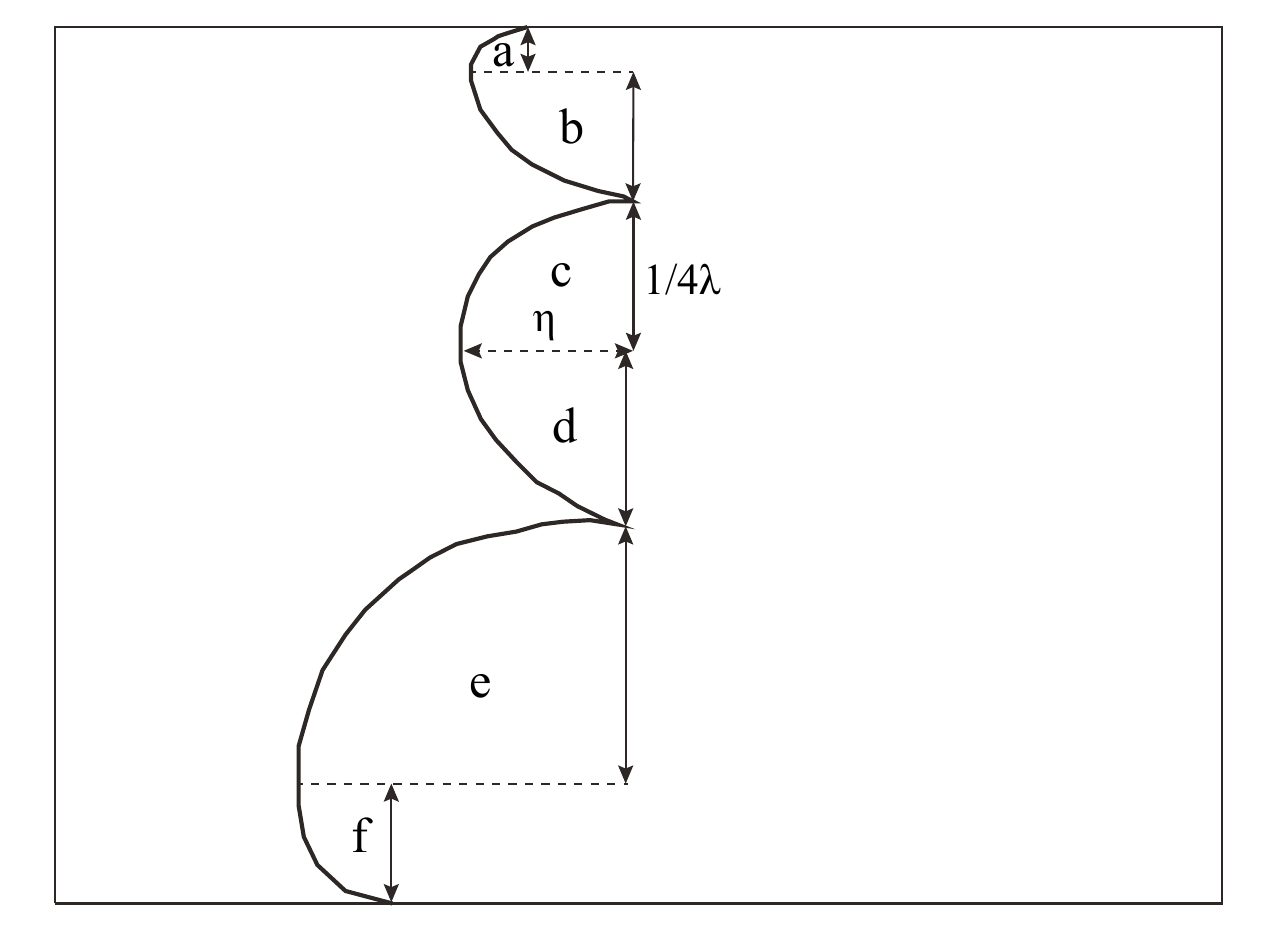}
	\caption{Sketch of the flame cells configuration. The amplitude and wavelength for cell a, b, c, d, e and f are measured separately.}
	\label{cell_sketch}
\end{figure}

Figure \ref{M1_9interaction} shows the detailed evolution of the interaction of a shock with propagation Mach number $M_\text{s}$ = 1.9 and a stoichiometric $\text{H}_{\text{2}}/\text{air}$ flame, which has an initial pressure of 17.2 kPa before ignition. The first two frames illustrate the cellular flame structure and evolution before the interaction. Note that comparing to the middle cell, the top and bottom cell are highly asymmetric, likely because of the top and bottom walls of the channel. The right propagating shock can be seen in the second frame to reach the flame interface. In the third frame, the incident shock wave (ISW) passed half the flame front, and gave rise to a curved transmitted shock wave (TSW) and a reflected expansion wave. Note that the incident shock reached the bottom cell first, and led to the earlier formation of the transmitted shock than the rest of the flame. Following the passage of the shock, the flame was flattened and pushed backwards to the burned gas. In the third and fourth frames, it is obvious that the flame developed two fronts with the passage of the incident shock. As shown in the sketch in figure\ \ref{3d_flame_sketch}, this is due to the fact that the flame was also initially curved in the third narrower dimension of our channel. With the passage of the shock, the 3D curved flame is subjected to the RM instability in both directions. The development of the flame front in the two directions thus leads to the two front features in the Schlieren images. The fourth to eighth frames show the further reversion of the flame cusps and the growth of the flame cells amplitudes. Here we define the amplitude to be the maximum horizontal distance from the tip to the root linking the intersection point of the cell with the adjacent cells, as marked as $\eta$ in figure\ \ref{cell_sketch}. With the development of the reversed flame, the formation of some roughness on the flame surface is evident.  As this was not observed in the 2D numerical simulations discussed below, this roughness may also be associated with 3D effects: irregular shock refractions in the third and/or interaction of the shock with the boundary layer flow, where very steep gradients of density are also present and may generate locally more vorticity than from the shock - cellular flame  interaction. With the elongation of the flame cusps, one can also observe the closing-up of the flame funnels, which implies the combustion toward the inside of the funnel. Note that the closing-up effects are more evident for smaller cells. Another point to notice is that although before the interaction the initial amplitudes of the cells are evidently different, the reversed funnels developed to lengths similar to each other in the last three frames.

\begin{figure}
	\centering
	\begin{subfigure}{0.7\columnwidth}
		\centering
		\includegraphics[width=1.0\linewidth]{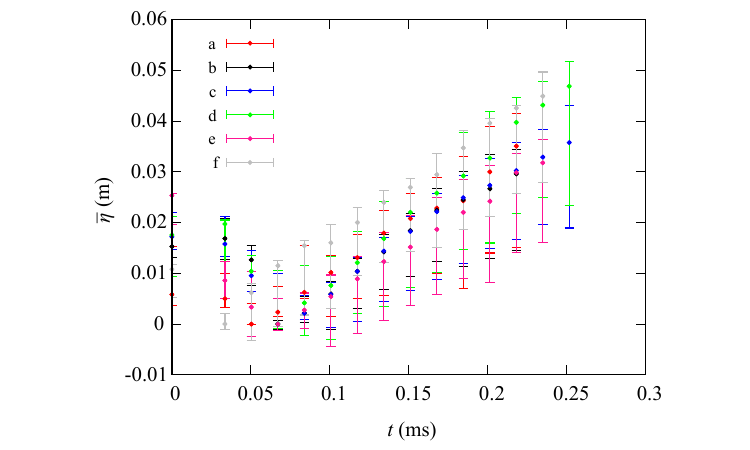}
		\caption{}
		\label{ms19_amplitude}
	\end{subfigure}
	\begin{subfigure}{0.7\columnwidth}
		\centering
		\includegraphics[width=1.0\linewidth]{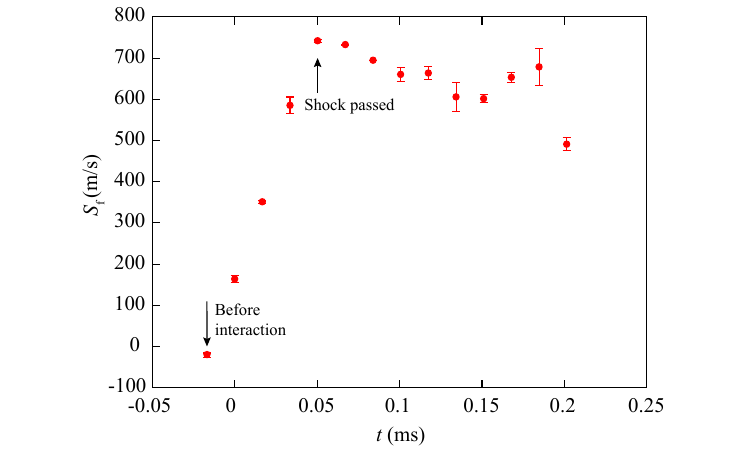}
		\caption{}
		\label{ms19_sf_exp}
	\end{subfigure}
	\caption{Time evolution of (a) normalized cell amplitudes and (b) average flame propagation speeds following the interaction of a $M_\text{s}$=1.9 incident shock wave with the stoichiometric hydrogen-air flame.}
	\label{ms19_sf_amplitude}
\end{figure}

To quantify the enhancement effect of the incident shock wave on the flame front, the cell amplitudes growth were measured and shown in figure\ \ref{ms19_amplitude}. Before the measurement of the cell amplitudes, an edge detection method was adopted to process the Schlieren images. As the experiment clearly indicates the 3D effects, during the development of the flame, there is also the wall effect which tends to slow down the flame, our measurement account for these uncertainties due to the superposition effects in the Schlieren images. In figure\ \ref{ms19_amplitude}, the points illustrate the leading edge of the flame front where, as illustrated in figure\ \ref{3d_flame_sketch}, the flame is uninfluenced by the wall. The positive and negative errors come from the 3D effects and the wall effects. An example of the measurement of the amplitudes and their uncertainties is given in Appendix A. Here we assume each half cell to be 1/4 of a sine wave, thus the wavelength can be defined as 4 times the vertical height of a half cell. In consideration of the cell asymmetry, as shown in the sketch in figure\ \ref{cell_sketch}, the amplitude was measured separately for the top and bottom half of each cell. To compare the amplitudes evolution of the different cells, we measured the amplitudes evolution in terms of the relative amplitude difference and normalized by the wavenumber. As mentioned in the MB theory, the average of the amplitudes before and after the interaction is a parameter that influences the growth rate of the amplitudes. The amplitudes were thus also divided by $({\eta_0}^{-}+{\eta_0}^{+})$ for comparison. Figure \ref{ms19_amplitude} shows the time evolution of the normalized amplitudes with the expression of $\overline{\eta} =\frac{\eta-{\eta_0}^{+}}{k({\eta_0}^{-}+{\eta_0}^{+})}$, where $\eta_0^{-}$ refers to the flame amplitude at time -0.03 ms and $\eta_0^{+}$ varies from 0.034 ms to 0.075 ms determined by the time at which the shock passed the according cell in this experiment. The error bar represents the uncertainty from the thick band in the measurement. As shown in figure  \ref{ms19_amplitude}, the shock first passed the cell (f) at approximately 0.034 ms and traversed all the cells at 0.075 ms, and led to the decrease of the flame amplitude. Subsequent to the interaction, the compressed flame interface grew nearly linearly with time for all the 6 half-cells before 0.15 ms, and then gradually become non-linear. Note the growth rate of the normalized amplitudes for all the cells are in good agreement with each other. Analysis of the sequential frames in the experiments also permits us to evaluate the average flame propagation speed following the interaction, as shown in figure\ \ref{ms19_sf_exp}. Assuming the flame is two-dimensional, the flame propagation speed is determined by the rate of change of the volume weighted leading edge flame position where the flame is uninfluenced by the wall. By taking the middle flange as a reference, we measured the volume between the middle flange to the flame front in each consecutive frames, the volume weighed distance was then evaluated by dividing the volume by the channel height and width. During the passage of the shock, the left moving flame was pushed back and acquired a maximum speed of 742 m/s. Then the propagation speed went through a decrease with fluctuation within the range of visualization after the shock has fully passed the flame, implying an increase in the burning velocity opposite to the propagation direction. The average propagation speed was used for the reconstruction of the flow field to evaluate the average burning velocity increase within the visualization time, as reported in Appendix B.  Over the time interval available in the experiments, the burning velocity was evaluated to be 7.1$\pm$4.5m/s for the case of the $M_s$=1.9 shock-flame interaction. 

\subsection{Effects of incident shock strength}
\begin{figure}
	\centering
	\includegraphics[width=0.8\linewidth]{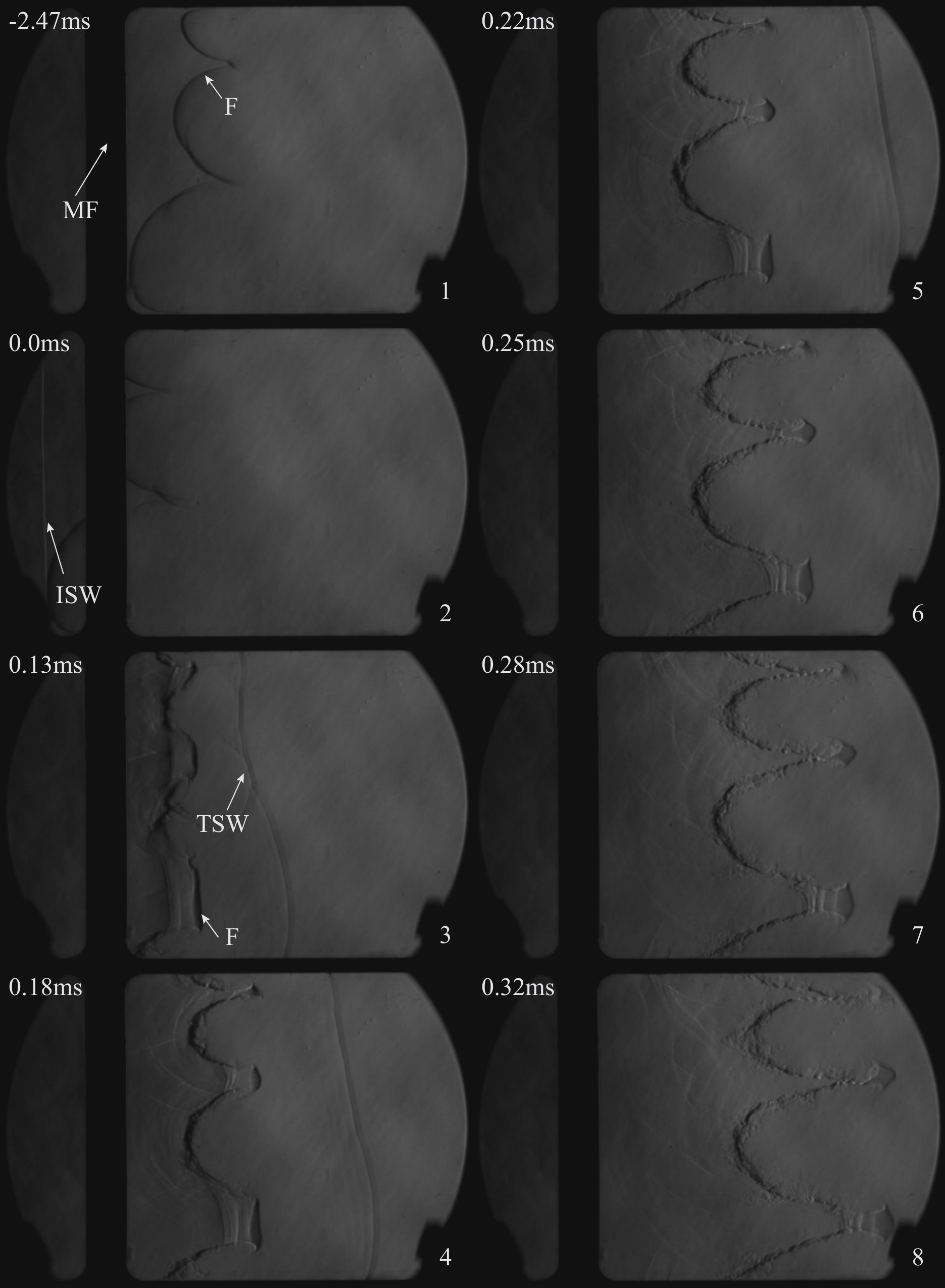}
	\caption{Schlieren image sequence of the interaction of a $M_s$ = 1.75 incident shock wave with stoichiometric hydrogen-air flame. Time 0 corresponds to the beginning of the interaction. Flame initial pressure of 17.2 kPa. Recorded at 59,590 fps.}
	\label{interaction_1_7psi}
\end{figure}

Figure\ \ref{interaction_1_7psi} shows the evolution of the flame with a weaker incident shock of Mach number 1.75. Similar to the case in figure\ \ref{M1_9interaction}, the first two frames show the structure of the cellular flame before the interaction. The shock reached the flame in the second frame, formed a transmitted shock wave and a rarefaction wave after passing the flame in the third frame. The fourth and fifth frames show the time interval between 0.18 ms and 0.22 ms after the interaction and are to be compared to the seventh and eighth frames in figure\ \ref{M1_9interaction}. One can notice that with lower incident shock strength, the flame reversion and propagation is slower than the case of the shock of $M_s$ = 1.9.  The flame amplitude also increased slower for the case with lower incident shock strength, while the closing up of the funnel shows no obvious difference between the two cases. The flame cusps further reversed in the sixth to eighth frames with larger amplitude burning rate toward the inside of the flame funnel. In the last frame, the very thin funnel neck implies that the mushroom caps were separating from the main structure. 

\begin{figure}
	\centering
	\begin{subfigure}{0.7\columnwidth}
		\centering
		\includegraphics[width=1.0\linewidth]{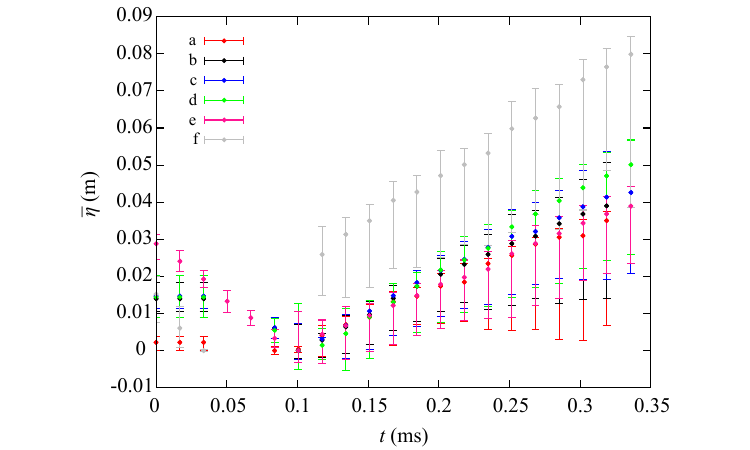}
		\caption{}
		\label{M175_amplitude}
	\end{subfigure}
	\begin{subfigure}{0.7\columnwidth}
		\centering
		\includegraphics[width=1.0\linewidth]{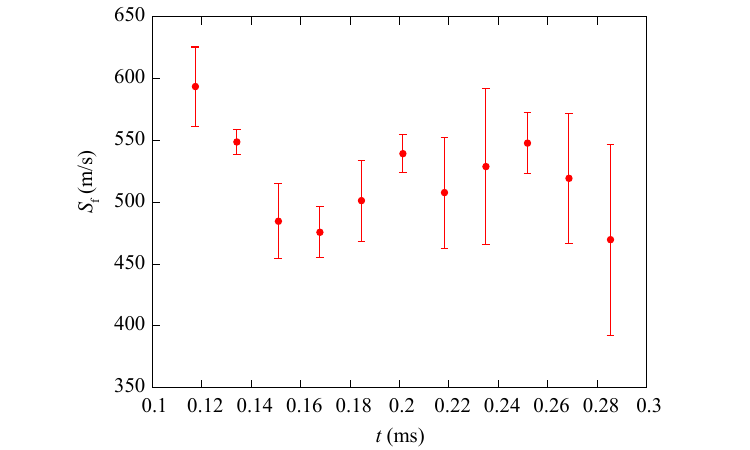}
		\caption{}
		\label{ms175_sf_exp}
	\end{subfigure}
	\caption{Time evolution of (a) normalized cell amplitude and (b) average flame propagation speed following the interaction of a $M_\text{s}$=1.75 incident shock wave with stoichiometric hydrogen-air flame.}
	\label{ms175_sf_amplitude}
\end{figure}

Figure \ref{ms175_sf_amplitude} shows the time evolution of the normalized amplitude for the case of $M_s =$ 1.75. Similarly, the shock passed the cell (f) before the rest of the cells. The amplitudes of the compressed flame interface, which formed during the passage of the incident shock, showed consistent nearly linear increase with time until approximately 0.2 ms for all the cells. Then nonlinear growth of the amplitudes can be observed from 0.2 ms. In the non-linear stage, the growth rate of the amplitude gradually decayed and slightly diverged between the different cells. The average flame propagation speed following the interaction is plotted in figure\ \ref{ms175_sf_exp}. Limited by the middle flange and the Schlieren system, we plotted only the flame displacement speed after the shock has passed. Subsequent to the interaction with the shock, the flame propagation speed gradually decreases with a fluctuation through the examined channel, which implies the increase in the burning velocity opposite to the flame propagation direction.

\begin{figure}
	\centering
	\includegraphics[width=0.8\linewidth]{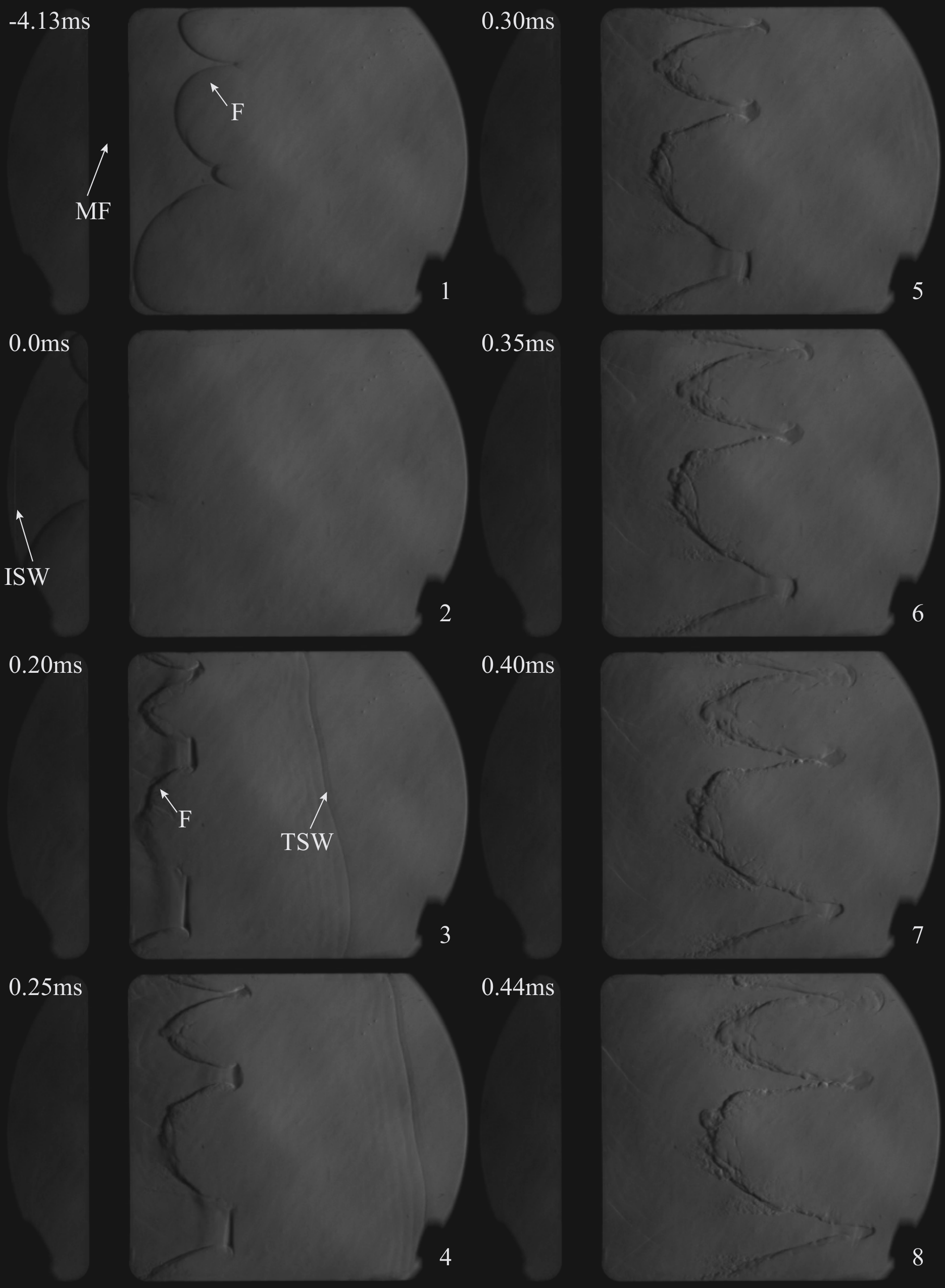}
	\caption{Schlieren image sequence of the interaction of a $M_\text{s}$ = 1.53 incident shock wave with stoichiometric hydrogen-air flame. Time 0 corresponds to the beginning of the interaction. Flame initial pressure of 17.2 kPa. Recorded at 59,590 fps. }
	\label{interaction_1_2psi}
\end{figure}

In figure \ref{interaction_1_2psi}, we present the interaction of the flame with further decreased incident shock Mach number of 1.53. The first stage of the interaction until 0.25 ms are similar to the results in figures\ \ref{M1_9interaction} and \ref{interaction_1_7psi}, although with moderate reversion, slower amplitude growth and more smooth flame front than the other cases due to the lower strength of the incident shock. As shown in the 5th to 8th frame, the flame tip gradually separated from the main structure 0.3 ms after the interaction because of the burn-out of the unburned gas inside the cell funnels.  

\begin{figure}
	\centering
	\begin{subfigure}[ht]{0.7\linewidth}
		\centering
		\includegraphics[width=1.0\linewidth]{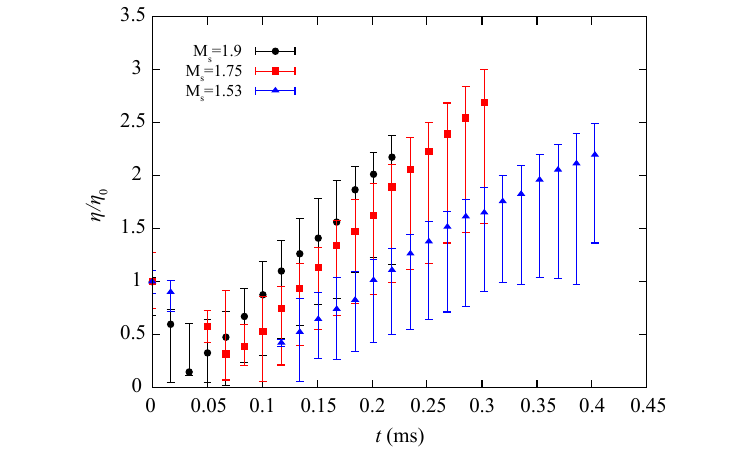}
		\subcaption{}
		\label{amplitude_exp_comp}
	\end{subfigure}
	\begin{subfigure}[ht]{0.7\linewidth}
		\centering
		\includegraphics[width=1.0\linewidth]{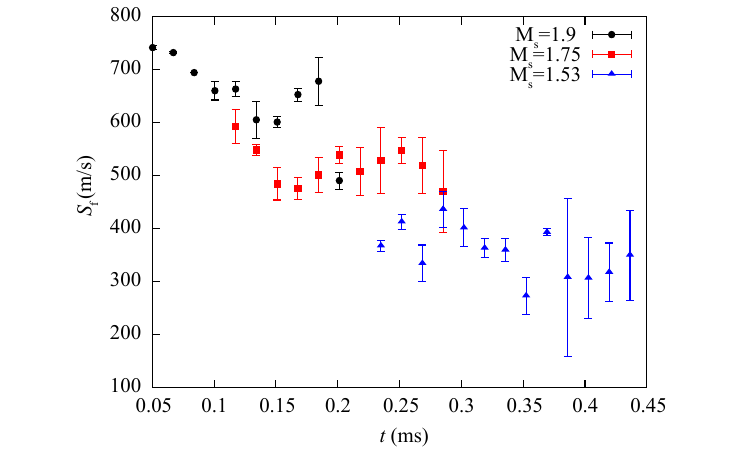}
		\subcaption{}
		\label{sf_exp_comp}
	\end{subfigure}
	\caption{Time evolution of (a) amplitudes of cell $d$ and (b) propagation speeds subsequent to the interaction for different incident shock strength of $M_\text{s}$ = 1.9, 1.75 and 1.53, $\eta_{0}$ corresponds to the respective initial amplitude before the interaction.}
	\label{amplitude-sf_exp_comp}
\end{figure}

Figure\ \ref{amplitude-sf_exp_comp} compared the evolution of cell amplitudes and flame propagation speeds after the shock fully passed the flame for all the incident Mach numbers considered. For simplicity, the middle-down cell was chosen for comparison. The trends were similar for all three cases considered. The growth of the amplitude and the flame propagation speed are more pronounced for higher Mach numbers. After the passage of the incident shock, the amplitudes grew nearly linearly. Then, as the flame evolved, the growth of the amplitude gradually slowed down. For the case with weak incident shocks, the decay of the growth rate is more obvious. 

\section{Numerical method and physical model}
\label{sec:numerical method and physical model}

In the experiments, the duration of the flame evolution is limited by the length of the apparatus and the dimension of the Schlieren visualization section. The evolution of the flame to later times was thus addressed numerically. We consider a two-dimensional problem in a rectangular domain, for which we model the flame evolution and interaction by the reactive Navier-Stokes equations. 

\begin{align}
	\frac{\partial \rho}{\partial t} + &\nabla \cdot (\rho \mathbf{u}) = 0,\\
	\frac{\partial \rho \mathbf{u}}{\partial t} + \nabla \cdot(&\rho \mathbf{u}  \mathbf{u})+\nabla p = \nabla \cdot \mathbf{\tau},\\
	\frac{\partial \rho e}{\partial t} + \nabla \cdot ((\rho e+p)\mathbf{u}) &= \nabla \cdot (\mathbf{u} \cdot \mathbf{\tau}) - \nabla \cdot (K \nabla T) + Q \dot \omega,\\
	\frac{\partial \rho Y}{\partial t} + \nabla \cdot (\rho &\mathbf{u} Y) + \nabla \cdot (\rho D \nabla Y) = \dot \omega,
\end{align}

\noindent where $\rho$, $\mathbf{u}$, \textit{p}, $e$ and \textit{Y} refer to mass density, velocity vector, pressure, specific total energy and mass fraction of the product, respectively. The equation of state satisfies the ideal gas law as follows:
\begin{equation}
	T = \frac{p}{R_{s} \rho}, \quad
	e=\frac{p}{\rho(\gamma -1)}+\frac{\mathbf{u}^2}{2},
\end{equation}

\noindent where $R_{s}$ is the specific gas constant. The reaction rate $\dot \omega$ is modeled by the single step Arrhenius kinetics,
\begin{equation}
\dot \omega = \rho A (1-Y) \exp(-\frac{E_a}{RT}).
\end{equation}

\noindent The viscous stress tensor is given by $\mathbf{\tau} = \mu (\nabla \mathbf{u} + (\nabla \mathbf{u})^{\text{T}} - \frac{2}{3}(\nabla \cdot \mathbf{u}) \mathbf{I})$, where $\mathbf{I}$ is the identity matrix. Other properties to note are the isentropic index $\gamma$, the viscosity $\mu$, the thermal conduction coefficient \textit{K}, the mass diffusion coefficient \textit{D}, the activation energy $E_{a}$, the pre-exponential factor \textit{A} and the universal gas constant $R_s$.   

The system of equations is non-dimensionalized as follows
\begin{equation}
	\left. \begin{array}{l}
		\displaystyle
	\tilde{\rho }=\frac{\rho }{\rho _0}, \quad    
	\tilde{u}=\frac{u}{S_L},\quad 
	\tilde{v}=\frac{v}{S_L},\quad 
	\tilde{P}=\frac{P}{\rho _0\left(S_L\right){}^2},\quad 
	\tilde{T}=\frac{\tilde{P}}{\tilde{\rho}}=\frac{R_sT}{\left(S_L\right){}^2},\\
	\displaystyle
	\tilde{e}=\frac{e}{\left(S_L\right){}^2},\quad 
	\tilde{x}=\frac{\rho _0 S_L c_p}{K} x,\quad 
	\tilde{y}=\frac{\rho _0 S_L c_p}{K} y,\quad 
	\tilde{t}=\frac{\rho _0 \left(S_L\right){}^2 c_p}{K} t,
\end{array} \right\}
\end{equation}
where the characteristic density $\rho_0$ is the initial density of the fresh gas, the characteristic speed $S_L$ is the one-dimensional laminar flame speed, the characteristic flame length scale $L_{fs}=\frac{K}{\rho _0 S_L c_p}$ gives a measure of the pre-heat zone thickness of steady planar flame \citep{sharpe2006nonlinear,strehlow1984combustion}, $t_{fs}=\frac{L_{fs}}{S_{L}}$ is the characteristic flame time. Here, $c_p$ denotes the specific heat at constant pressure. Furthermore, it is useful to mention the non-dimensional Lewis ($L_e$) and Prandtl ($P_r$) numbers:
\begin{equation}
	L_e = \frac{K}{\rho c_p D}, \quad
	P_r = \frac{\mu c_p}{K}.
\end{equation}

To model the stoichiometric hydrogen-air mixture at 17.2 kPa initial pressure and 294 K initial temperature, the fresh gas properties and the heat release were chosen to fit the laminar free flame. The input parameters for the model are given in table\ \ref{table_input}. In table\ \ref{table_input}, the initial flame density, isentropic index, fluid viscosity, and heat conductivity were evaluated for the unburned gas composition using Cantera thermal-chemical tools and the Li mechanism \citep{li2004updated}, and the isentropic index, fluid viscosity, and heat conductivity were kept constant through out the simulation. The laminar flame speed and the flame thickness $L_f$ were calculated from the one-dimensional free flame. The heat release was evaluated by the enthalpy difference between the fresh and burned gases. The Lewis number was set to be 1.0 following \citet{jomaas2007transition}. The global activation energy was determined from 
\begin{equation}
	E_{a} = -2R \left( \frac{\partial ( \ln (\rho_{0} S_{L}))}{\partial (\nicefrac{1}{T_{ad}})} \right)_{p_0,\phi}
\end{equation}
following \citet{egolfopoulos1990chain}, by calculating the mass burning rate ($\rho_{0} S_{L}$) for the given initial pressure($p_0$) and equivalence ratio ($\phi$), then slightly varying its value through the substitution of a small quantity of nitrogen by argon. Here, $T_{ad}$ is the adiabatic flame temperature. 

\begin{table}
	\begin{center}
		\def~{\hphantom{0}}
		\begin{tabular}{lccccccc}
			$\rho_0$ & 0.1474 kg/m$^{3}$ & $S_L$ & 1.98 m/s& $\gamma$ & 1.4013 & $L_f$ & 0.0034 m  \\
			$K/(\rho c_p)$ & 2.6 $\times 10^{-4}m^2$/s & $\mu$ & 1.8 $\times 10^{-5}$ pa s & $E_a/R$ & 27390 K$^{-1}$ & $Q$ & 3.0 $\times 10^{6}$ J/kg \\
			\textit{A} & 4.43 $\times 10^{11} s^{-1}$ & $P_r$ & 0.4655 & $L_e$ & 1.0 
		\end{tabular}
		\caption{Thermo-chemical properties and model parameters for the stoichiometric hydrogen-air combustion at T=294 K and $p_0$ = 17.24 kPa.}
		\label{table_input}
	\end{center}
\end{table}

\begin{figure}
	\centering
	\begin{subfigure}{0.495\columnwidth}
		\centering
		\includegraphics[trim={1.0cm 0cm 1cm 0cm},clip,width=1.0
		\columnwidth]{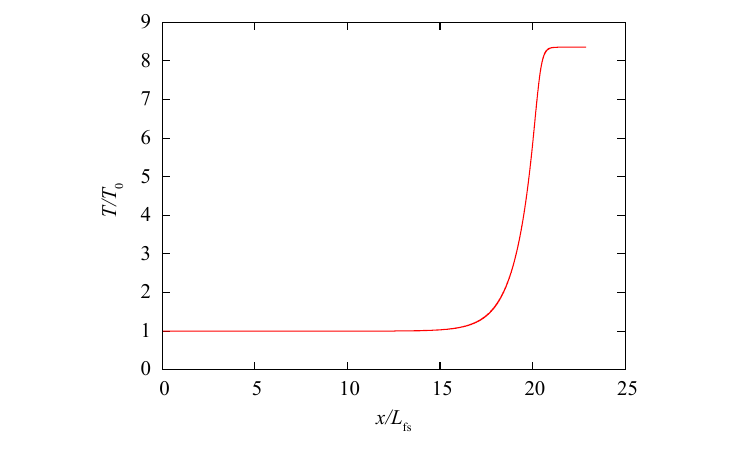}
		\subcaption{}
		\label{T}
	\end{subfigure}
	\hfill
	\begin{subfigure}{0.495\columnwidth}
		\centering
		\includegraphics[trim={1.0cm 0cm 1cm 0cm},clip,width=1.0
		\columnwidth]{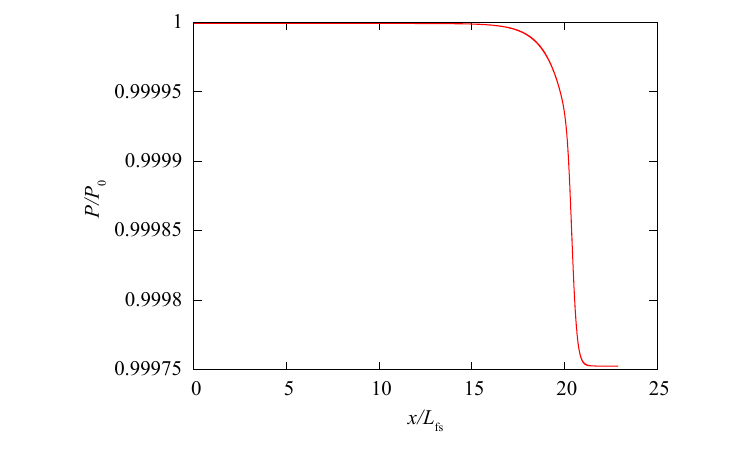}
		\subcaption{}
		\label{P}
	\end{subfigure}
	\begin{subfigure}{0.495\columnwidth}
		\centering
		\includegraphics[trim={1.0cm 0cm 1cm 0cm},clip,width=1.0
		\columnwidth]{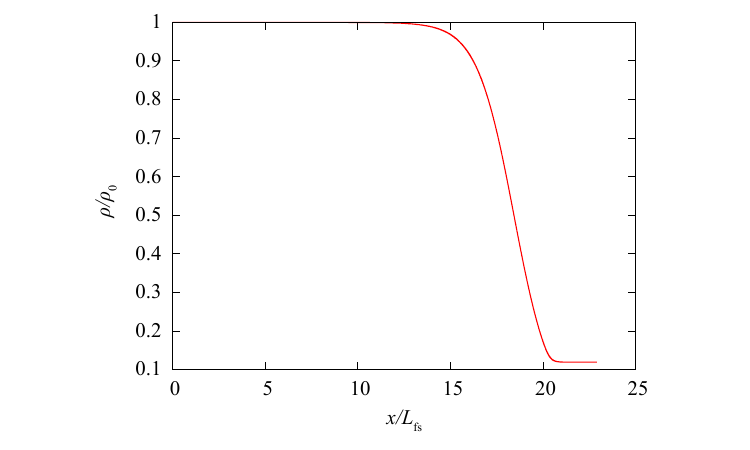}
		\subcaption{}
		\label{rho}
	\end{subfigure}
	\hfill
	\begin{subfigure}{0.495\columnwidth}
		\centering
		\includegraphics[trim={1.0cm 0cm 1cm 0cm},clip,width=1.0
		\columnwidth]{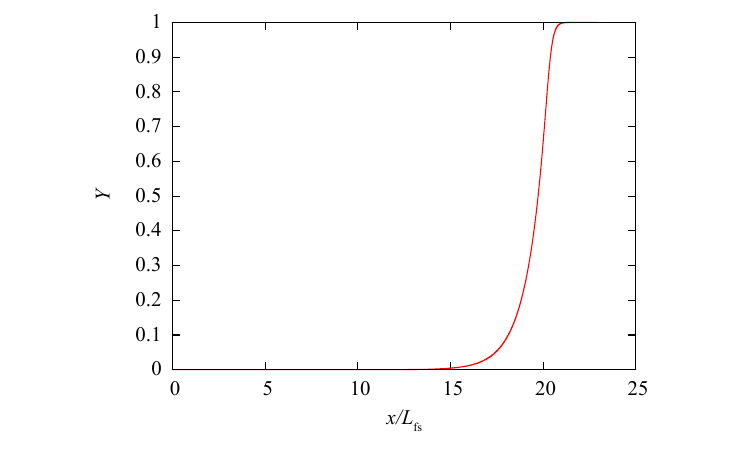}
		\subcaption{}
		\label{Y}
	\end{subfigure}
	\caption{Steady 1D flame profile of temperature, pressure, density and reaction progress variable.}
	\label{1d_flame_profile}
\end{figure}

The pre-exponential factor was calculated by solving a system of equations for the laminar flame speed satisfying the compressible one-dimensional steady flame structure using the shooting method proposed by \citet{travnikov1997stability}. Figure\ \ref{1d_flame_profile} shows the results of the structure for temperature, pressure, density and reaction progress variable of the 1D steady flame in the frame of reference the flame. Note that the pressure only changes very little, given that the flame burning velocity is much smaller than the sound speed in the fresh gases. The flame thickness, which was determined as $(T_b-T_0)/(\frac{dT}{dx})_{max}$, was evaluated to be 1.48$L_{fs}$ and later used to scale to the laminar flame thickness evaluated from the experiments. To fit the experiments, we also define $t_{f}=\frac{L_{f}}{S_{L}}$ to be the laminar flame time.

The cellular flame before the interaction was developed by imposing a small perturbation on the evaluated 1D flame in the longitudinal direction. The perturbed flame was placed in the middle of a rectangular channel with a domain length of 1000 flame lengths and a height of 16.5 $L_{fs}$ that represents the width of the cell (c) in figure\ \ref{cell_sketch} or the dashed rectangle in figure\ \ref{M1_9interaction} (2) before the interaction. The perturbation was set to be $A_0 \cos(2\pi y)/\lambda$ where $A_0=0.1 L_{fs}$. The flame then freely evolved to form a steady cellular structure, which helps to prevent to deal with the initial unsteady evolution effect from subsequent dynamics. When the flame has propagated sufficient enough to form the steady cellular flame structure, the jumps in the fluid state parameters corresponding to a shock with a Mach number in the range of 1.53 to 2.5 were imposed in the left, 50 flame lengths from the flame, with the post shock parameters determined from the Rankine-Hugoniot relations. The length to the right of the flame was then varied within a range between 500 and 2000 $L_{fs}$ to investigate the flame evolution after the interaction. The top and bottom boundaries were set to be symmetric while the left and right boundaries had imposed zero-gradient conditions.

The numerical model detailed above was solved using a finite-volume code developed by S.\ Falle at the University of Leeds \citep{maxwell2018modelling} with second-order accurate Godunov Riemann solver and adaptive mesh refinement \citep{falle1993numerical}. The method has been demonstrated in the papers of \citet{sharpe2006nonlinear} and \cite{sharpe2008numerical} for the simulation of flame propagation using the Navier-Stokes equations. In the simulation, a base grid of 3 points per flame thickness with five refinement levels was used. The reaction and diffusion zone were enforced to have the highest refinement level, giving an effective resolution of 48 points per flame thickness, which is higher than that suggested in \citet{sharpe2006nonlinear} for properly resolving a cellular flame described by one-step Arrhenius kinetics. The refinement also ensured that all shock waves formed by the interaction were properly captured. The results of resolution studies are given in Appendix C.

\section{Numerical results}
\label{sec:numerical results}

\subsection{Establishment of a cellular flame}
\begin{figure}
	\centering
	\begin{subfigure}[ht]{0.49\linewidth}
		\centering
		\includegraphics[trim={1.2cm 0cm 1cm 0cm},clip,width=1.0\columnwidth]{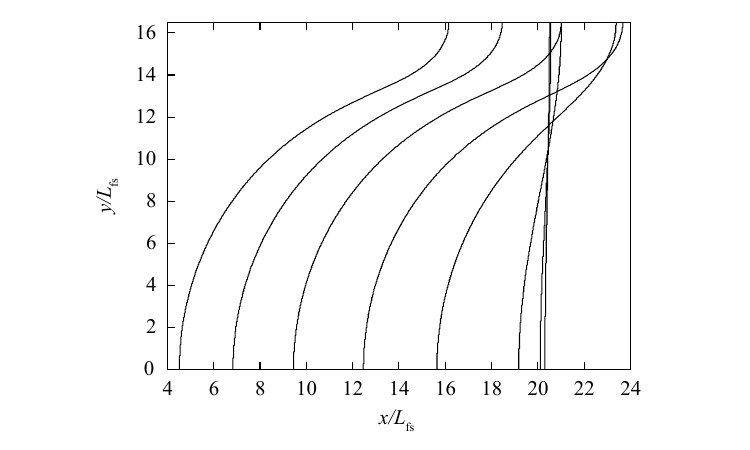}
		\subcaption{}
		\label{sim_flame}
	\end{subfigure}
	\hfill
	\begin{subfigure}[ht]{0.495\linewidth}
		\centering
		\includegraphics[trim={0.8cm 0cm 1cm 0cm},clip,width=1.0\columnwidth]{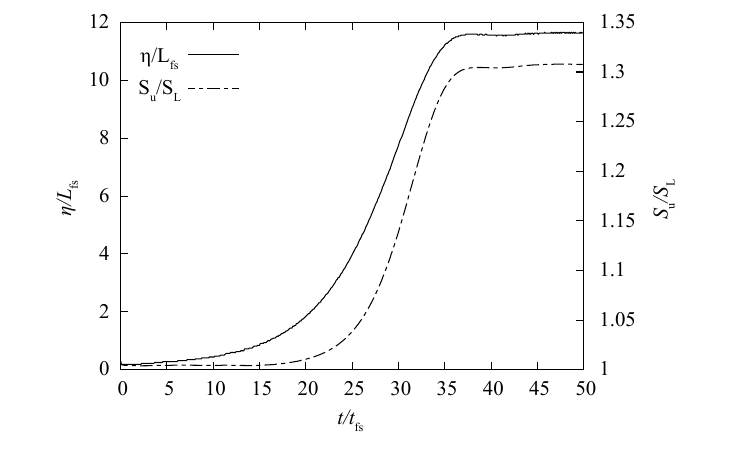}
		\subcaption{}
		\label{steady_su_amp}
	\end{subfigure}
	\caption{Time evolution of: (a) maximum heat release contour at times $t/t_{fs}$ = 0, 10, 20, 30, 35, 40, 45 and 49.9 illustrating the initial setup and the cellular flame development, (b) flame shape amplitude and burning velocity.}
	\label{2d_steady_cellular_flame}
\end{figure}

Figure \ref{2d_steady_cellular_flame} shows the evolution of an initially planar flame, to which an initial perturbation of half cosine wave with an amplitude of $0.01 L_{fs}$ was applied. The flame contour at successive times was obtained from the locus of maximum heat release along the flame.  Following the initial perturbation, the flame deformed and saturated at a quasi-steady cellular structure after approximately 35 $t_{fs}$.   The evolution of the flame shape amplitude shown in Fig.\ \ref{steady_su_amp}, defined as before as the largest $x$-direction distance on the flame front, shows the characteristic exponential growth of the flame shape.  The same growth is evident in Fig.\ \ref{steady_su_amp} for the flame burning velocity, defined as 
\begin{equation}
S_{u} = \frac{\iint_{s} \dot \omega dxdy}{H \bar{\rho}_{u}}, 
\end{equation}
where $\bar{\rho}_{u}$ is the average density of the unburned gas in the section of the channel occupied by the flame, $H$ is the domain height. 

\begin{figure}
	\centering
	\includegraphics[width=0.4\linewidth]{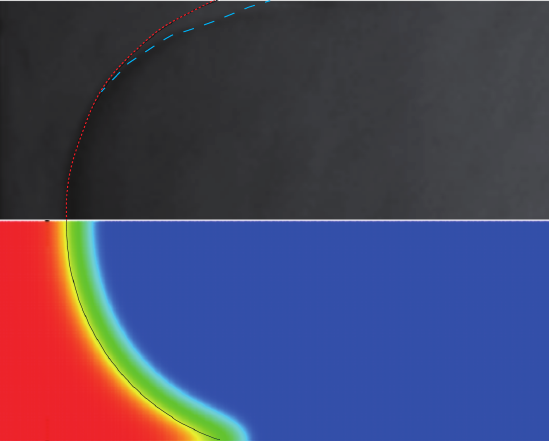}
	\caption{Comparison of the flame shape from the top-half middle cell marked in figure\ \ref{M1_9interaction} (2) before the interaction (top) and the simulation (bottom), the red dashed line and the black line denote the profile of the simulation, the blue dashed line shows the difference.}
	\label{flame_shape_comp}
\end{figure}

The steady cellular flame acquired a burning velocity of 1.3$S_L$ and an amplitude of 11.6$L_{fs}$. The steady structure recorded at $t=49.9t_{fs}$ was then compared with the reference cell from the experiment, as shown in figure\ \ref{flame_shape_comp}. Note that although the flame structure in the simulation is achieved by a perturbation development, the acquired structure is in very good agreement with the experimentally determined shape, with some differences observed at the cell corners. Once such a steady flame was developed, shocks with varying Mach numbers were generated 50 flame lengths from the flame at time 49.9$t_{fs}$ to have the interaction.  

\subsection{Interaction of $M_\text{s}$=1.9 shock with flame}

\begin{figure}
	\centering
	\includegraphics[width=1.0\linewidth]{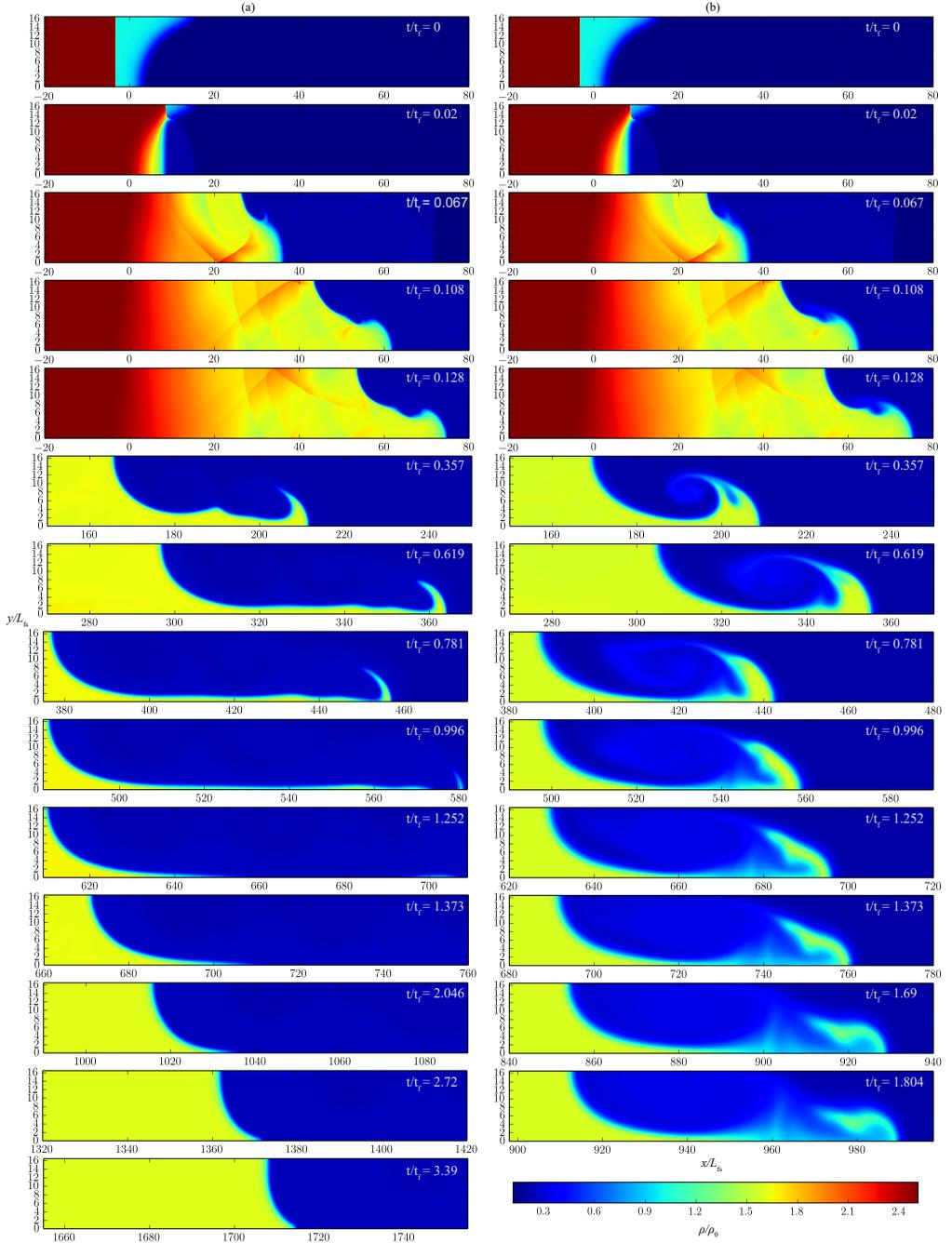}
	\caption{Density profiles illustrating the (a) flame and (b) inert interface evolution subsequent to the interaction with the $M_\text{s}$=1.9 shock.}
	\label{sim_M19sfi}
\end{figure}

Figure \ref{sim_M19sfi}a illustrates the detailed evolution of the interaction of the cellular flame and a shock with an initial Mach number of 1.9. The first frame shows the incoming shock and the flame interface before the interaction. Similar to the experimental results, the interface was flattened (2nd frame) and pushed backwards (3rd frame to the last) to the burned gas. In the 3rd frame, the incident shock wave traversed the interface and formed a transmitted shock wave, a rarefaction wave and a series of transverse waves. A small bump can be observed in the middle of the interface. As the funnel shaped interface further reversed, the small bump punctured down and pushed the interface tip to propagated further, more transverse waves can be observed as the interface propagated from the 4th to 5th frames. The first 5 frames show the same trend of flame flattened and reversed back as the experiment in figure\ \ref{M1_9interaction}. As the flame funnel stretched further in the 6th to 9th frames, it progressively developed into a very long and narrow neck that is close to the bottom boundary. In the 10th frame, the flame funnel burned out the fresh gas along its neck and left only the cellular root. The remained flame front then gradually decayed to another cellular flame, as shown in the 11th to 14th frame.

\begin{figure}
	\centering
	\includegraphics[width=1.0\linewidth]{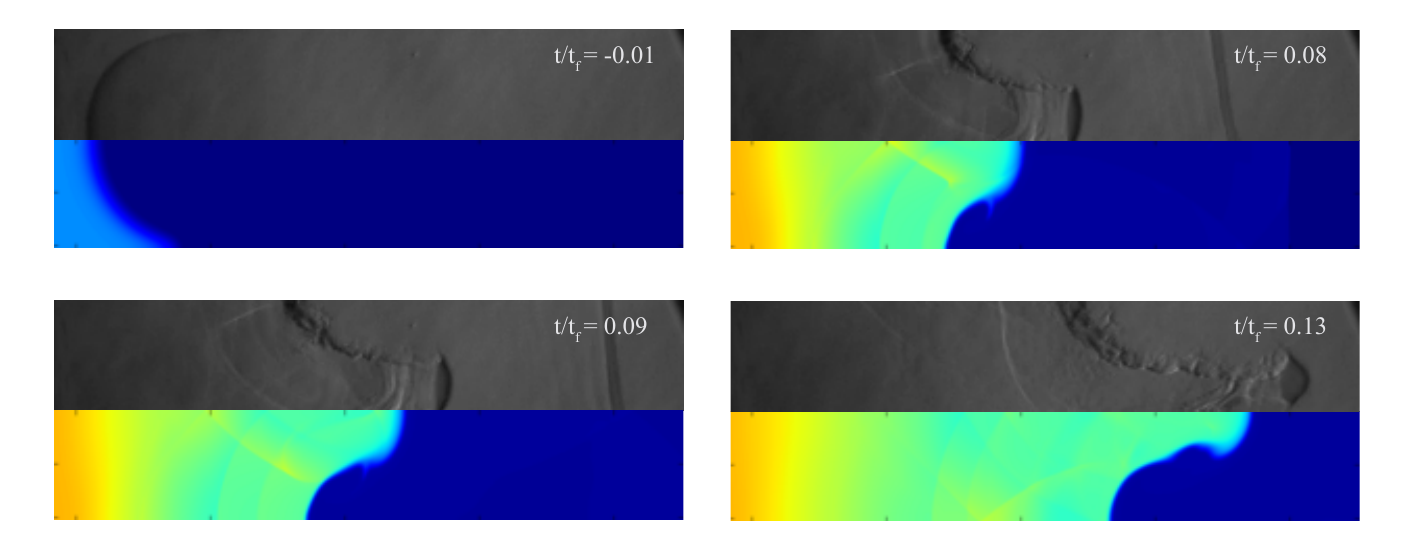}
	\caption{Density profiles illustrating the time evolution of the flame following the interaction with the $M_\text{s}$=1.9 shock from the simulation and the experiment.}
	\label{sim_M19sfi_comp_exp}
\end{figure}

Figure\ \ref{sim_M19sfi_comp_exp} compares the time evolution of the flame following the interaction with the $M_\text{s}$=1.9 shock from the simulation with the experiment. A general good agreement of the deformation dynamics is obtained, while some dissimilarities are evident.  Firstly, the amplitude of the flame is somewhat larger in the experiments.  This may due to the larger initial amplitude of the experimental cell, since the RM instability proportionally amplifies the initial disturbance amplitude. Secondly, the numerically calculated 2D deformation provides further interpretation of the double front observed experimementally, which was attributed to 3D effects.  The line of sight integration in the experimental visualization does not discriminate between deformations near walls and in the center of the channel.  Finally, the simulations do not capture all the small scale features observed in the experiments.  Our grid convergence study did not reveal changes at finer resolutions, even for resolutions up to 100 points per flame thickness.   The disturbances observed in the experiments are likely due to 3D effects involving deformation of the flame and reverberating shocks in the third dimension, and notable influence of boundary layers as a locus of vorticity deposition from the interaction with the shock.  The increase in the surface area of the flame in the experiments may be responsible for the higher growth in the experiments.  Clearly, further 3D simulations incorporating effects not addressed here, such as more realistic chemistry and its potentially involving smaller scale shock-chemistry interactions absent in the 1 step model used, and inclusion of wall effects, which typically also involve condensation phenomena in the boundary layers \citep{boust2012unsteady}, would be required to clarify this issue; this is however outside the scope of the present study, which addresses the primary mode of deformation cellular flames after interacting with a shock.

To further investigate the influence of the inert RM instability on the shock-flame interaction, the reaction was artificially turned off right before the shock reached the flame, as shown in the first frame in figure\ \ref{sim_M19sfi}b. The first 5 frames show the same trend of the interface flattened and reversed back as the reactive simulation and the experiment. In the 6th frame, the concave structure formed by the initial small bump merged with the interface tip and further evolved to a spike topped off with a mushroom cap, implying the classic non-linear RM instability structure. Unlike the smooth flame front of the reactive case, the development of the vortex at the tip of the mushroom cap can be observed in the 6th to the last frame. During the interface propagation, the growing vortex dissipated the top of the interface. In the 10th to 13th frame, the growth of the interface become less obvious. 

\begin{figure}
	\centering
	\begin{subfigure}[ht]{0.495\linewidth}
		\centering
		\includegraphics[trim={1.5cm 0.3cm 1cm 0cm},clip,width=1.0\linewidth]{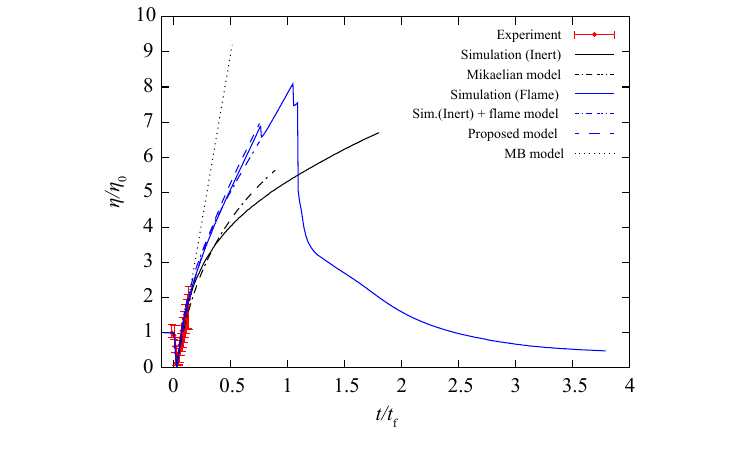}
		\subcaption{}
		\label{ms19_sim_amp}
	\end{subfigure}
	\hfill
	\begin{subfigure}[ht]{0.495\linewidth}
		\centering
		\includegraphics[trim={1.5cm 0.3cm 1cm 0cm},clip,width=1.0\linewidth]{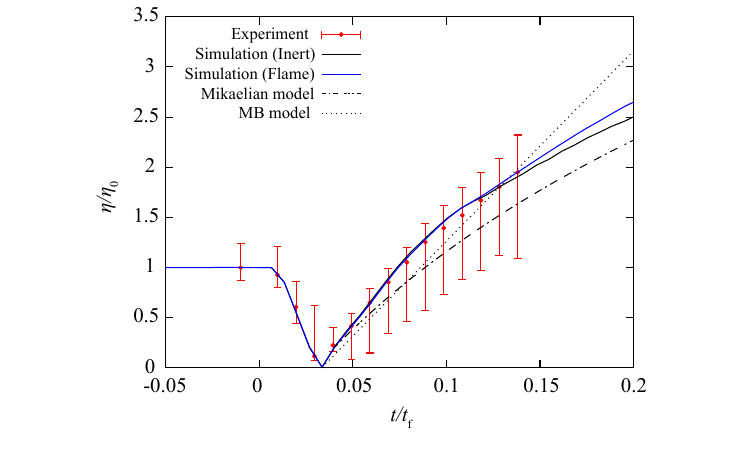}
		\caption{}
		\label{ms19_sim_amp_zoom}
	\end{subfigure}
	\caption{(a) Comparison of the interface amplitude as a function of time for the experiment, the simulation of the reactive and non-reactive interaction, and the prediction of Mikaelian's model, combined model of the inert simulation and the flame model, and the proposed model, (b) zoomed-in view of interface amplitude for the experiment, the simulation of the reactive and non-reactive interaction, and the prediction of the Mikaelian model and the MB model within the time range of 0 to 0.2 $t_f$ subsequent to the interaction with the Mach 1.9 incident shock.}
	\label{ms19_sim_amp_ampzoom}
\end{figure}

Figure\ \ref{ms19_sim_amp} illustrates the amplitude development of the interface from the reactive and inert simulation of the RM instability. To track the interface evolution in the simulation, we imposed a scalar \textit{Y} in the flow field assuming \textit{Y} = 0.0 and 1.0 to be the heavier and lower density region and \textit{Y} = 0.0 to 1.0 along the interface. The amplitudes were then acquired by tracking the contour of the scalar value of Y = 0.2. For comparison, the time evolution of the flame amplitude for the experiment of the interaction of the same incident shock Mach number is also plotted here. By looking at figure\ \ref{ms19_sim_amp}, we can conclude that the flame amplitude went through four distinct stages after the reversion of the interface. The first stage can be recognized as the inert evolution of the interface according to the inert RM instability, before approximately 0.12$t_f$. Evident from the zoomed-in figure of the amplitude evolution until 0.15$t_f$ in figure\ \ref{ms19_sim_amp_zoom} is that the reactive and the inert simulations are nearly identical from 0 to 0.12$t_f$. Moreover, the amplitude evolution from the simulations is in excellent agreement with the experiment. In the second stage, the volumetric expansion caused by the combustion along the flame front gradually became important and resulted in a longer amplitude comparing to the inert case up to 1 laminar flame time ($t_f$). In this stage, the amplitude growth rate gradually decreased and became smaller than the first stage. Note that the small drop at 0.77$t_f$ and 1.05$t_f$ are due to the separation of the mushroom cap from the main structure. The third stage started from 1.07$t_f$ when the flame consumed the fresh gas inside the funnel and gave rise to an abrupt collapse of amplitude. In the last stage, the remaining flame spike went through a gradual decay of amplitude to reach the new steady state in the post shock flow subjected to the Landau-Darrieus instability.

\begin{figure}
	\centering
	\includegraphics[trim={1.5cm 0.3cm 1cm 0cm},clip,width=0.70\linewidth]{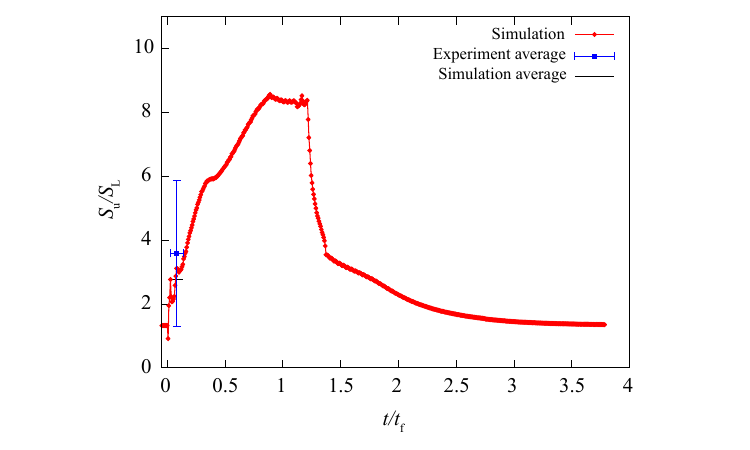}
	\caption{Time evolution of the flame burning velocity after the interaction of a Mach 1.9 incident shock.}
	\label{ms19_sim_su}
\end{figure}

Figure \ref{ms19_sim_su} shows the evolution of the flame burning velocity and a short time average value corresponding to the observation time in the experiments.  The passage of the incident shock quickly increased the burning velocity to approximately twice the initial value, then it gradually increased with the growth of flame area in the inert evolution stage. Within the experimental visualization time, the burning velocity increase is about 2.76 times the laminar flame speed. In the reactive evolution stage, the burning velocity developed with a smaller rate and went through a change with an even lower increase rate at 0.5$t_f$ when it formed the long neck along the bottom line. It peaked at approximately 0.75 laminar flame time while the mushroom cap started to separate from the main spike, then developed with mild decrease at about 8.5 times the laminar burning velocity until the flame finally burned all the fresh gases along the funnel neck, demonstrated by the sharp decrease in the burning velocity at about 1 laminar flame time. In the last stage, the burning velocity gradually asymptoted to a value close to the flame velocity before the interaction.  

The average burning velocity in the simulations is found to be in agreement with the experiment, within the uncertainties of the model used to infer $S_u$ described in Appendix B.  The experimental value was found 1.3 times larger, possibly suggesting 3D effects in the experiments, but this difference remains within the uncertainty of the method. 

\subsection{Influence of the incident shock strength}

\begin{figure}
	\centering
	\includegraphics[width=1.0\linewidth]{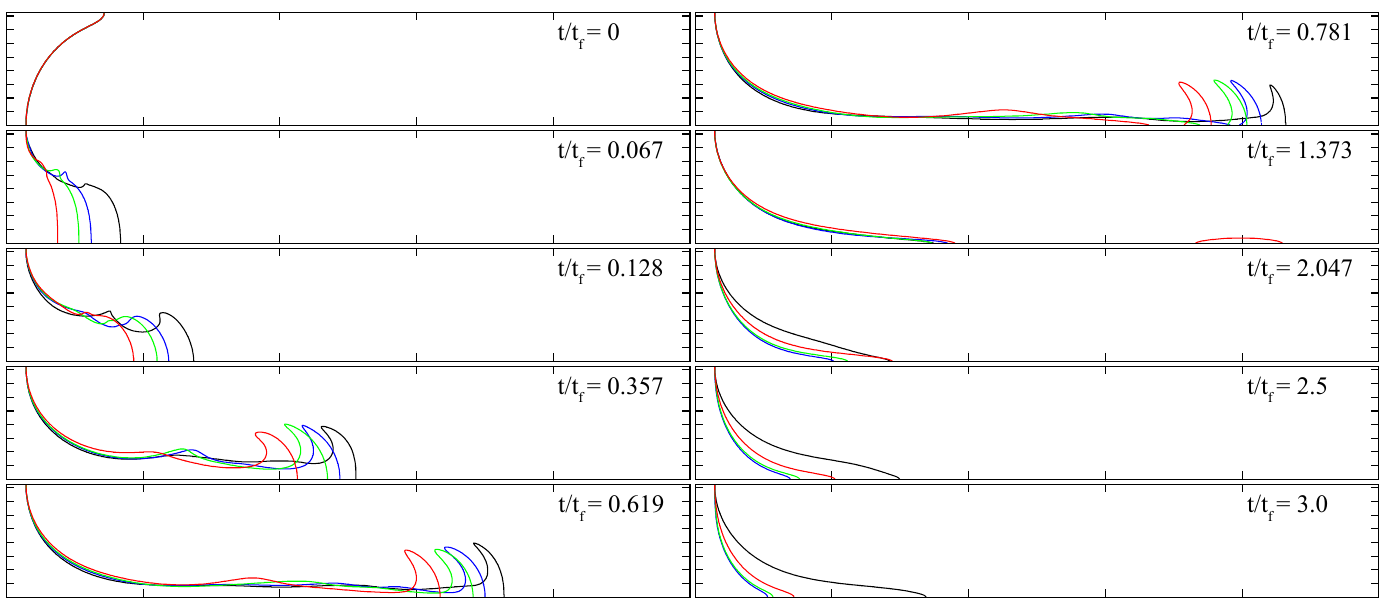}
	\caption{Time evolution of the superimposed flame contour of $Y$ = 0.2 for the interaction with different incident shock strengths illustrated with the same size as of figure\ \ref{sim_M19sfi}. The red, green, blue and black lines refer to the cases of shock Mach number of 1.53, 1.75, 1.9 and 2.5.}
	\label{comp_sim_contour}
\end{figure}

In the experiments, the highest shock Mach number of 1.9 was adopted to avoid exceeding the maximum allowable operating pressure of the facility for safety concerns. The interaction of the flame with higher incident shock Mach number of 2.5 was thus addressed in the simulation. The effect of shock strength is illustrated in figure \ref{comp_sim_contour} where the evolution of the flame contours was obtained numerically as explained above for shock Mach numbers of 1.53, 1.75, 1.9 and 2.5. For comparison, we rearranged and superimposed the flame contour of $Y$ = 0.2 for cases with various incident shock strength at each time step such that they all start from the same coordinate. It is evident that the overall flame deformation followed the same trend of interface reversion, funnel elongation and area increase, mushroom-cap separation and burn-out of the funnel neck, and new stage cellular flame development. With the increase in shock Mach number, the development of flame front area and shape amplitude are more pronounced. Also, it can be noted that, the flame at the new cellular flame development stage tends to become thinner for the cases with higher incident shock Mach numbers.

\begin{figure}
	\centering
	\includegraphics[trim={1.5cm 0.3cm 1.0cm 0cm},clip,width=0.65\linewidth]{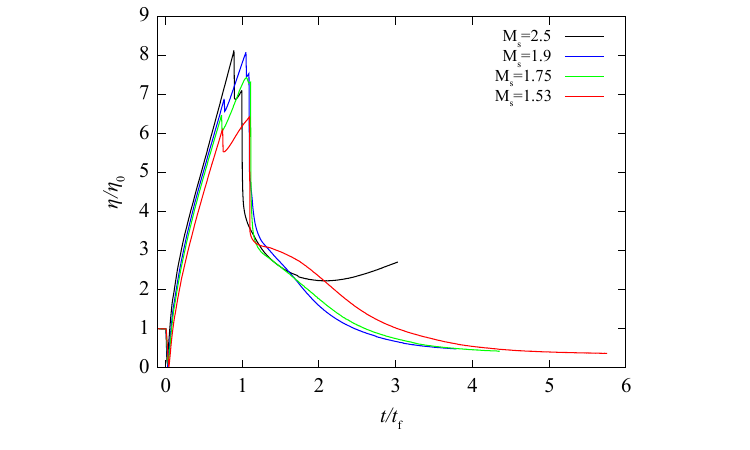}
	\caption{Comparison of the time evolution of amplitudes of the flame subsequent to the interaction with the different incident shock Mach number of 1.53, 1.75, 1.9 and 2.5.}
	\label{comp_sim_amp}
\end{figure}

\begin{figure}
	\centering
	\begin{subfigure}[ht]{0.495\linewidth}
		\centering
		\includegraphics[trim={1.5cm 0.3cm 1.0cm 0cm},clip,width=1.0\linewidth]{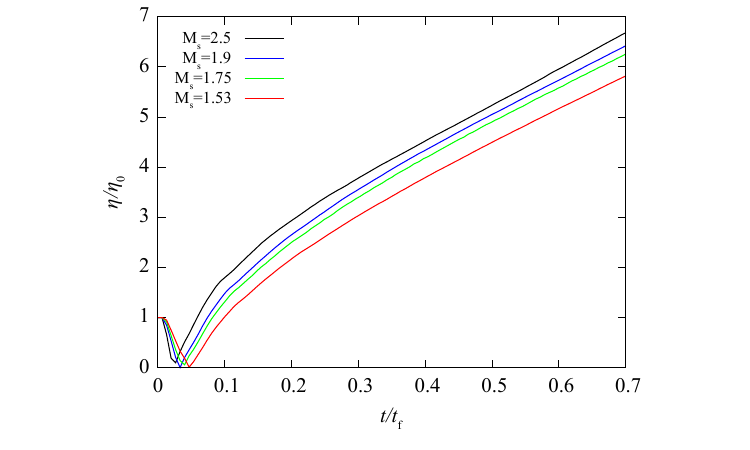}
		\subcaption{}
		\label{comp_sim_amp_s}
	\end{subfigure}
	\hfill
	\begin{subfigure}[ht]{0.495\linewidth}
		\centering
		\includegraphics[trim={1.5cm 0.3cm 1.0cm 0cm},clip,width=1.0\linewidth]{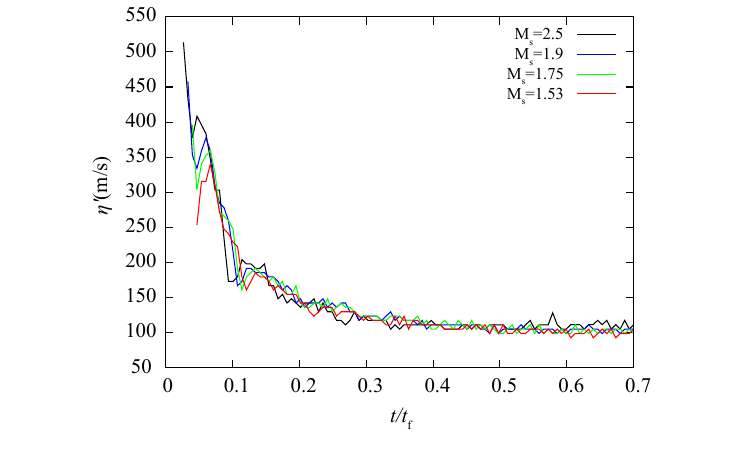}
		\subcaption{}
		\label{comp_sim_ampgrs}
	\end{subfigure}
	\caption{Zoomed-in view of (a) the time evolution of the flame amplitudes from 0 to 0.7$t_f$ and (b) their growth rate in the amplification stages for the interaction with the different incident shock Mach number of 1.53, 1.75, 1.9 and 2.5.}
	\label{comp_sim_amps_grs}
\end{figure}

\begin{figure}
	\centering
	\includegraphics[trim={1.5cm 0.3cm 1.0cm 0cm},clip,width=0.65\linewidth]{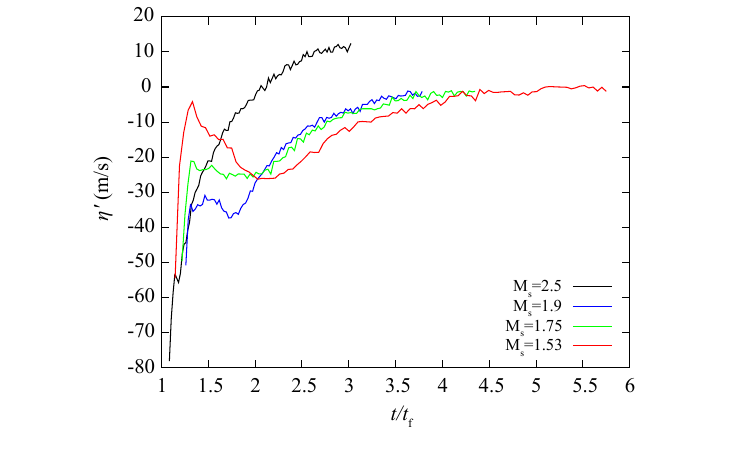}
	\caption{Comparison of the time evolution of amplitudes growth of the flame in the new cellular flame formation stage for the cases of different incident shock Mach numbers of 1.53, 1.75, 1.9 and 2.5.}
	\label{comp_sim_amps4}
\end{figure}

We then examined the flame shape amplitudes and their growth rate for all the cases considered to determine the effect of shock strength on the flame front deformation. As shown in figure\ \ref{comp_sim_amp}, the overall evolution were qualitatively similar for the various shock Mach numbers investigated.   Larger flame shape amplitudes were achieved with stronger shocks.   The flame amplitude went through the same trend of decrease during the traverse of the incident shock, and increase due to the RM instability and the chemical reaction. Hereafter, the amplitudes went through a drastic decrease due to the burn-out of the fresh gas along the flame funnel neck, and then gradually decay to reach the new steady state for the case of incident Mach number of 1.53, 1.75 and 1.9, while the amplitude of the flame that interacted with the $M_s$=2.5 shock gradually increased. To examine the flame evolution with more detail, figure \ref{comp_sim_amps_grs} shows the enlarged view of the amplitudes and growth rates evolution at early times before the collapse of the interface for the cases considered. As shown in figure\ \ref{comp_sim_amps_grs}, the flame interfaces were compressed to the minimum values faster for higher incident Mach numbers. The initial post-shock amplitude growth toward the burnt gas also showed the same tendency with the increase of Mach number at time before 0.075 $t_f$. Also, as shown in figure\ \ref{comp_sim_amp_s}, the growth rate mildly fluctuated within 0.075 $t_f$ for small Mach numbers comparing to the larger Mach numbers. Surprisingly, the growth rate of all cases considered decayed with the rate close to each other from approximately 0.075 $t_f$ until when the mushroom-cap started to disconnect from the main spike. The sharp decrease of amplitude for example at 0.73 $t_f$ for the case of $M_s$=1.53 implies the first separation of the flame tip. Then amplitude of the remained structure kept growing until the flame burnt out the gas along the funnel neck as shown in figure\ \ref{sim_M19sfi} (10th to 11th frame) at approximately 1 $t_f$ for all the cases, denoted by the sharp decrease of amplitudes starting from the case of higher Mach numbers. The fact that the amplitude growth is weakly influenced by the incident Mach number might suggest that the later flame development is not solely controlled by an inert RM instability. At last, as demonstrated in figure\ \ref{comp_sim_amps4}, the flame amplitudes all decayed and asymptoted to a value about half the initial state while the rates asymptoted to approximately zero for the cases of $M_s=$1.53, 1.75 and 1.9. The only exception is the $M_s$=2.5 case, where the decay rate of the flame amplitude gradually passed 0 and asymptoted to 12 m/s over the time of the calculation.  The saturation of the amplitudes growth is indicative of the saturation to a new cellular structure controlled Landau-Darrieus instability, while the non-steadyness observed at $M_s = $2.5 is attributed to the typical chaotic dynamics of cellular flames \citep{sivashinsky1983instabilities}, For the stronger shock case, the post shock flow reaches higher pressures and temperatures and the flame thickness is reduced.  For example, the flame thickness at the post shock state for the $M_s = $2.5 case is approximately half the flame thickness for the case of $M_s = $1.9.   With the higher compression of the gas ahead resulting in thinner flames , the flames evolve in an essentially larger channel.  It is thus not unreasonable to expect non-linear effects of cells forming and disappearing to become predominant in the cellular dynamics, such as those illustrated in our experiments in Fig.\ \ref{flame_later}.

\begin{figure}
		\centering
		\includegraphics[trim={1.5cm 0.3cm 1cm 0cm},clip,width=0.7\linewidth]{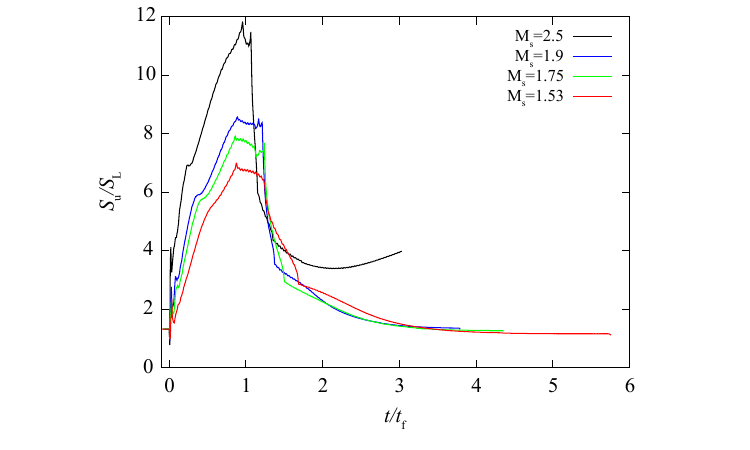}
		\caption{Evolution of the the flame burning velocity following the interaction with the different incident shock Mach number of 1.53, 1.75, 1.9 and 2.5.}
	\label{comp_sf_su}
\end{figure}

The differences in flame shape evolution correlate very well with the evolution of the burning velocity shown in figure\ \ref{comp_sf_su}.   The development of all the flame propagation speeds and burning velocities are qualitatively consistent to the aforementioned case of $M_\text{s}$=1.9, although the flame propagated with larger speed for the case with stronger incident Mach numbers. 

\section{Discussion}
\label{sec:discussions}
\subsection{Four evolution stages of the flame deformation}

The results presented in figures\ \ref{sim_M19sfi}, \ref{ms19_sim_amp_ampzoom}, and \ref{comp_sim_amp} suggest there are four distinct evolution stages of the flame amplitude after the passage of the shock and flame front inversion.  The overall flame evolution follows the trend of inert evolution stage dictated by the non-linear Ricthmyer-Meshkov instability, the reactive growth stage where volumetric expansion due to energy release enhances the growth of the inert RM instability, the flame funnel collapse stage and the new cellular flame development stage.  These are discussed in more detail below and a novel model for the reactively assisted RM instability is formulated. 

\subsubsection{Inert evolution stage}
\label{1Dreconstruction}

The first stage of the flame evolution after the flame is flattened by the incident shock can be recognized as the inert evolution of the interface, at times significantly less than the characteristic flame time, i.e., smaller than 0.12$t_f$ for the case with the incident Mach number of 1.9, as shown in figures\ \ref{ms19_sim_amp_ampzoom} and \ref{comp_sim_amp_s}. In this stage, the growth rate of the flame interface is controlled by the inert RM instability. 

When the interface amplitude is sufficiently small, the growth of the amplitude can be predicted by the linear model \citep{richtmyer1960taylor,meyer1972numerical}. For our case of shock passing from a fresh gas (heavy fluid) to a burned gas (light fluid), \citep{meyer1972numerical} proposed the following expression (MB model) for the initial linear growth rate: 
\begin{equation}
\frac{d\eta}{dt} = \frac{\eta_0^{+}+\eta_0^{-}}{2}k[u]A_t^{+},
\end{equation}
where $\eta$ is the amplitude of the perturbation, $\eta_0^{-}$ and $\eta_0^{+}$ are the amplitudes before and immediately after the shock passage, and $A_t^{+}$ the Atwood number after the interaction. As the amplitude develops to be comparable to or larger than the wavelength, the growth of the amplitude becomes nonlinear. \citet{mikaelian2003explicit} suggested a model that joined the linear and non-linear regime for predicting the amplitude growth rate for arbitrary Atwood numbers. The Mikaelian model gives:
\begin{equation}
\eta_{Mik}' =  \frac{\eta_0'}{1+3 \eta_0' k t \frac{(1+A_t)}{(3+A_t)}},  
\end{equation}  
where $\eta_0'= \frac{1}{2}(\eta_0^{-} +\eta_0^{+}) [u] k A_t^{+}$ is the amplitude growth rate in the linear stage \citep{meyer1972numerical}. 

In figures\ \ref{ms19_sim_amp_ampzoom} and \ref{comp_model_amp_ampgr}, we compared the experiment and numerical simulations results with the MB and the Mikaelian models. As shown in figure\ \ref{ms19_sim_amp},the MB model shows good agreement with the initial amplitude evolution for the inert interface, while the interface evolution after the shock-inert interface interaction can be well approximate by the Mikaelian model. Relatively good agreement can also be observed for the two models in the inert evolution stage in figure \ref{comp_model_amp_ampgr}, the discrepancy between the model and the experimental and simulation results might caused by the finite thickness of the interface rather than the discontinuity. It is obvious that for stronger incident shock Mach numbers, the inert evolution stage lasts for longer times. Also, the Mikaelian model shows better performance at later times for the case with the higher incident shock Mach number.

An interesting aspect of these results is that the shock-flame experiment provides a simple and promising method to further study the RM instability.  For observation times an order of magnitude less than the flame consumption time, the interface remains essentially inert.  This method provides a unique opportunity to study sharp interfaces devoid of separating membranes or non-desired mixing from injected gas curtains or sliding valves.    

\subsubsection{Reactive growth stage}
As illustrated in figure\ \ref{ms19_sim_amp_ampzoom}, when the time is comparable with the characteristic flame time $t_f$, the growth rate of the flame shape  amplitude can no longer be captured by the inert RM instability dynamics.  The gases processed by the flame expands and occupies a non-negligible volume affecting the motion of the flame shape.  

A model for the growth rate of the flame accounting for this supplementary volume expansion of the material crossing the flame, which lengthens the flame and increases its surface area can be formulated as a linear correction to the inert non-linear growth rate model:
\begin{equation}
	\frac{d \eta}{dt} =  \frac{d \eta_{rmi}}{dt} + \frac{d \eta_q}{dt},  \label{eq:combo}
\end{equation} 
where we define $\frac{d \eta}{dt}$, $\frac{d \eta_{rmi}}{dt}$ and $\frac{d \eta_q}{dt}$ to be the total flame interface growth rate, the growth rate caused by the inert RM instability and the heat release. Here, we expect the correction to be independent of the inert model.

\begin{figure}
	\centering
	\includegraphics[width=0.7\linewidth]{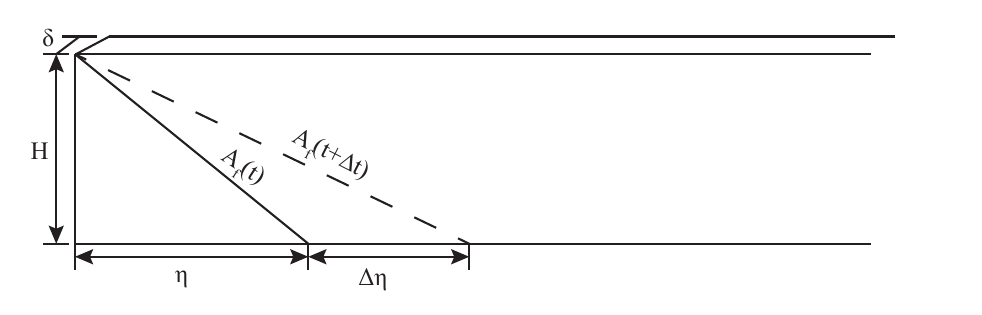}
	\caption{Simplified geometrical model of flame front expansion.}
	\label{model_illustration}
\end{figure}

We assume a flame front with a simple structure of a plane area, as shown schematically in figure\ \ref{model_illustration}. According to the mass conservation, the rate of volume increase of the burned gas is equal to the mass consumption rate of the fresh gas during the development of the flame:
\begin{equation}
	\frac{d}{dt} (\rho_b \frac{H}{2} \eta_q \delta) = \rho_u A_f S_u \delta,  
	\label{flame_struc0}
\end{equation}
where $A_f=\sqrt {\eta_q^2+H^2}$ is the flame area, $H$ is the channel height, $\delta$ is the channel width. Equation\ \ref{flame_struc0} thus can be re-written as:
\begin{equation}
	\frac{d}{dt} (\rho_b \frac{H}{2} \eta_q) = \rho_u \sqrt {\eta_q^2+H^2} S_u, \label{flame_struc} \\
\end{equation}

Equation\ \ref{flame_struc} implicitly gives the growth rate of the flame amplitude. It is thus integrated to yield
\begin{equation}
	\eta_q = \frac{1}{2}(\eta_{q0}+\sqrt {\eta_{q0}^2+H^2}) exp(\frac{2 \rho_u S_u}{\rho_b H} t)-\frac{H^2}{2(\eta_{q0}+\sqrt {\eta_{q0}^2+H^2})}exp(-\frac{2 \rho_u S_u}{\rho_b H} t). \label{flame_eta} 
\end{equation}
The constant of integration $\eta_{q0}$ is the initial flame amplitude. Hence, the growth rate of the flame length (and surface area) depends on the surface of the flame itself, yielding an exponential feedback. 
The reactive correction to be applied in expression \eqref{eq:combo} is the time derivative of \eqref{flame_eta}, $\frac{d \eta_q}{dt}$. 

The first test of the model proposed is to combine the result obtained for $\frac{d \eta_q}{dt}$ with the numerical result for the inert evolution shown in figure \ref{ms19_sim_amp}.  The result, also shown in figure \ref{ms19_sim_amp}, labeled "Sim. (Inert) + flame model" recovers very well the reactive simulation.

\begin{figure}
	\centering
	\begin{subfigure}[ht]{0.495\linewidth}
		\centering
		\includegraphics[trim={1.5cm 0.3cm 1cm 0cm},clip,width=0.95\linewidth]{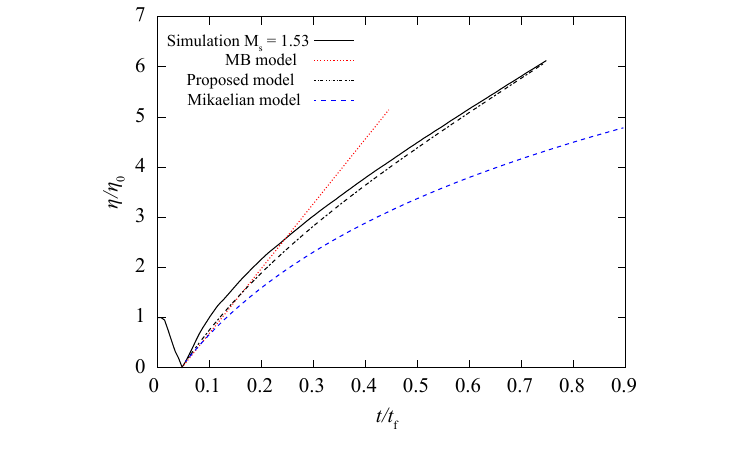}
		\subcaption{}
		\label{ms153_amp}
	\end{subfigure}
	\hfill
	\begin{subfigure}[ht]{0.495\linewidth}
		\centering
		\includegraphics[trim={1.5cm 0.3cm 1cm 0cm},clip,width=0.95\linewidth]{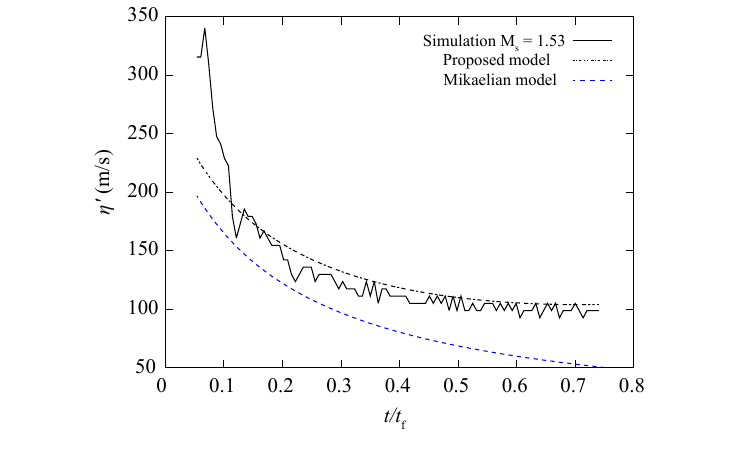}
		\subcaption{}
		\label{ms153_ampgr}
	\end{subfigure}
	\begin{subfigure}[ht]{0.495\linewidth}
		\centering
		\includegraphics[trim={1.5cm 0.3cm 1cm 0cm},clip,width=0.95\linewidth]{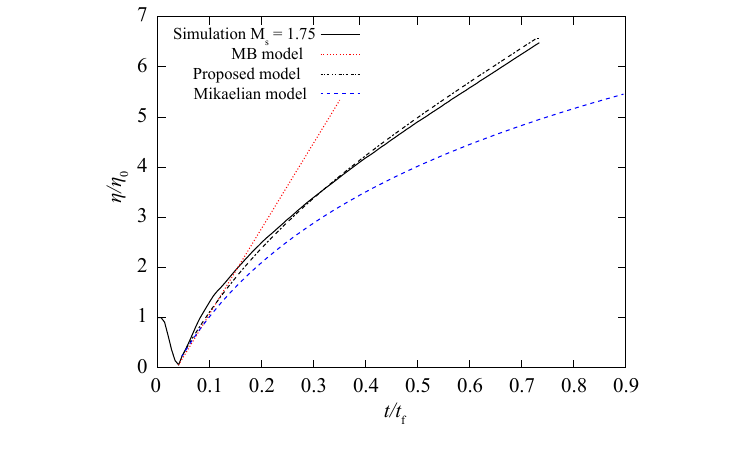}
		\subcaption{}
		\label{ms175_amp}
	\end{subfigure}
	\hfill
	\begin{subfigure}[ht]{0.495\linewidth}
		\centering
		\includegraphics[trim={1.5cm 0.3cm 1cm 0cm},clip,width=0.95\linewidth]{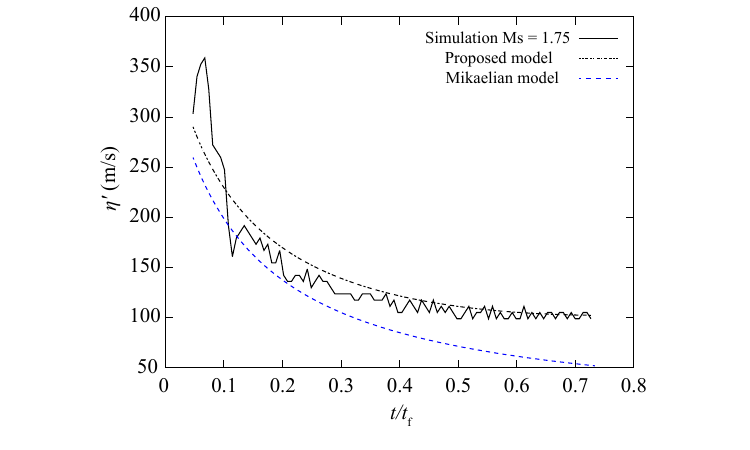}
		\subcaption{}
		\label{ms175_ampgr}
	\end{subfigure}
	\begin{subfigure}[ht]{0.495\linewidth}
		\centering
		\includegraphics[trim={1.5cm 0.3cm 1cm 0cm},clip,width=0.95\linewidth]{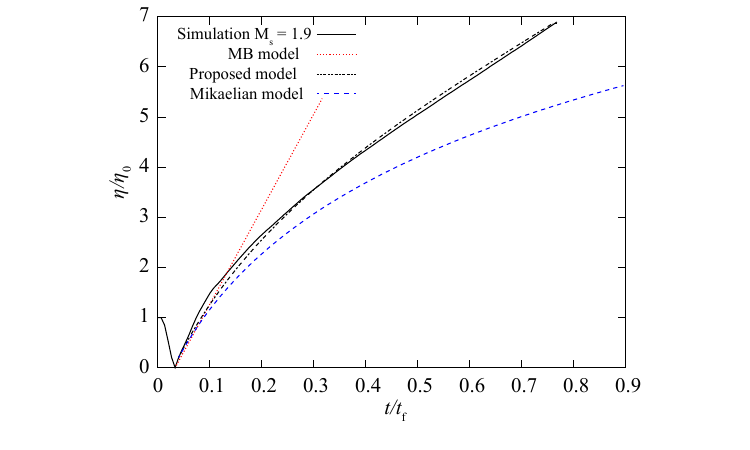}
		\subcaption{}
		\label{ms19_amp}
	\end{subfigure}
	\hfill
	\begin{subfigure}[ht]{0.495\linewidth}
		\centering
		\includegraphics[trim={1.5cm 0.3cm 1cm 0cm},clip,width=0.95\linewidth]{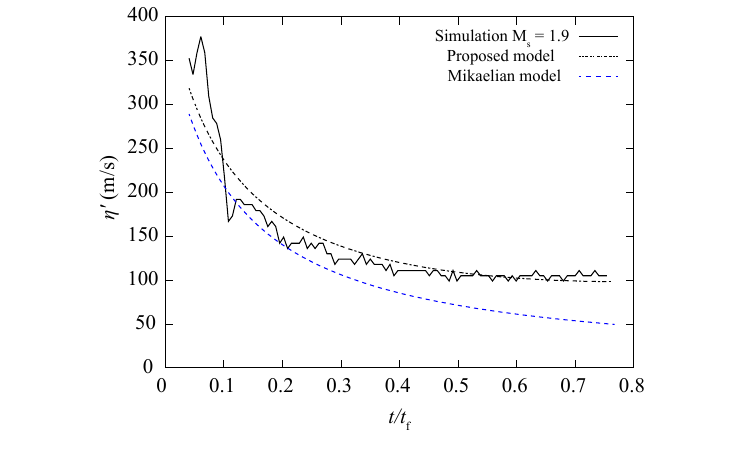}
		\subcaption{}
		\label{ms19_ampgr}
	\end{subfigure}
	\begin{subfigure}[ht]{0.495\linewidth}
		\centering
		\includegraphics[trim={1.5cm 0.3cm 1cm 0cm},clip,width=0.95\linewidth]{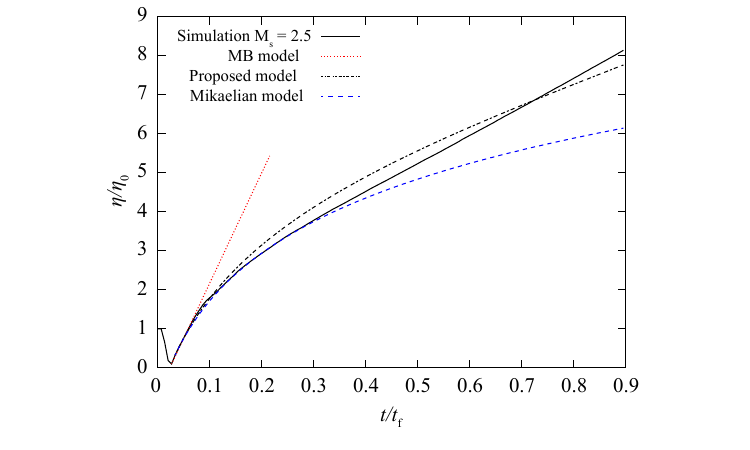}
		\subcaption{}
		\label{ms25_amp}
	\end{subfigure}
	\hfill
	\begin{subfigure}[ht]{0.495\linewidth}
		\centering
		\includegraphics[trim={1.5cm 0.3cm 1cm 0cm},clip,width=0.95\linewidth]{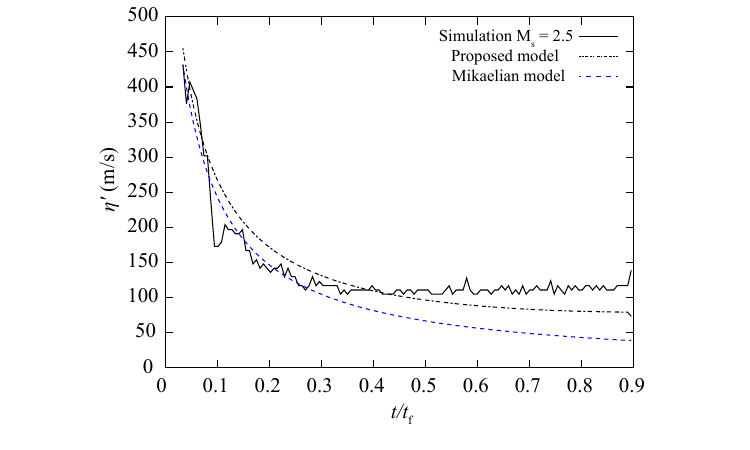}
		\subcaption{}
		\label{ms25_ampgr}
	\end{subfigure}
	\caption{Comparison of the time evolution of amplitude and growth rate of the simulation with proposed model for the interaction of flame with different incident shock Mach number of 1.53, 1.75, 1.9 and 2.5.}
	\label{comp_model_amp_ampgr}
\end{figure}  

A closed form expression for the reactive RM instability can be written using \eqref{eq:combo}, the correction $\frac{d \eta_q}{dt}$ and an algebraic model for the inert RM instability.  Since the Mikaelian model can well predict the non-reactive evolution of the inert interface, as shown in figure\ \ref{ms19_sim_amp} and \ref{comp_model_amp_ampgr}, the growth rate in the second stage can be modeled as:
\begin{equation}
	\frac{d \eta}{dt} = \eta_{Mik}' + \frac{d \eta_q}{dt}. 
	\label{equ1}
\end{equation}
The time evolution of the flame amplitude predicted by the proposed model was then compared with the result from the simulation and the combined model of the inert RMI interface evolution with the flame expansion model for $M_s$=1.9 shock interacts with the cellular flame, excellent agreement can be observed in figure\ \ref{ms19_sim_amp}. 

We then verified the proposed model by the comparison of the amplitude and its growth rate evolution with the simulation results of the interaction of the flame with different incident shock strengths. As shown in figure\ \ref{comp_model_amp_ampgr}, the overall flame evolution can be very well predicted by the proposed model. Note that for stronger incident shocks, the inert RM instability plays dominant roles for longer time following the shock-flame interaction, i.e. the inert Mikaelian model is in good agreement with the simulation until 0.3$t_f$ for the case of $M_s$=2.5 incident shock flame interaction. While for weaker incident shocks, the chemical reaction of the flame starts to play important roles earlier. Note that the chemical reaction seems to start to play roles only 0.05$t_f$ after the passage of the shock $M_s$=1.53, whereas the inert RM instability controlled the flame evolution for more than 0.25$t_f$ after the $M_s$=2.5 shock. Thus, the inert RM instability plays dominant roles following the shock-flame interaction. 

\subsubsection{Progressive decay of smaller scales}
\begin{figure}
	\centering
	\includegraphics[width=0.8\linewidth]{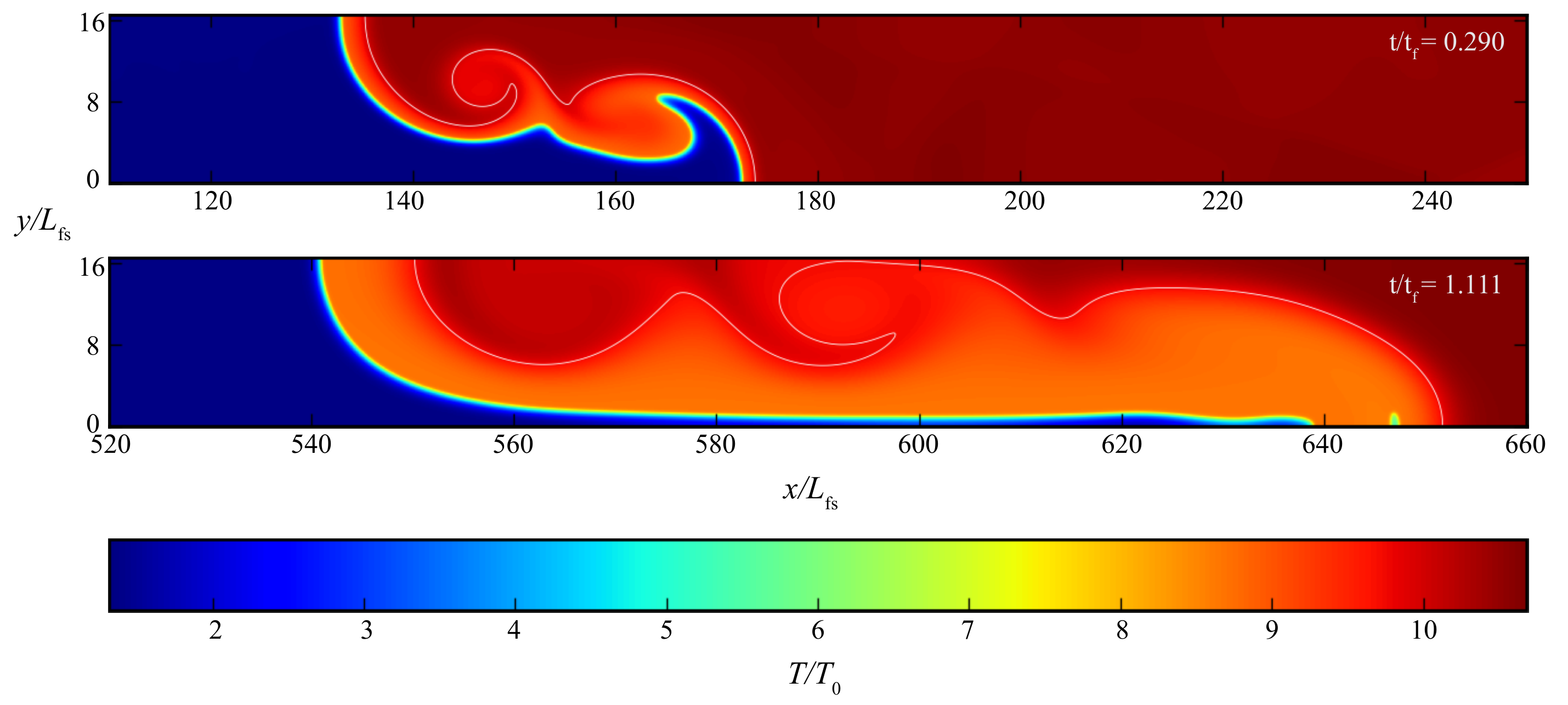}
	\caption{Temperature profile illustrating the flame and contact surface evolution subsequent to the interaction with the $M_\text{s}$ = 1.9 shock. The white curve denotes the position of the contact surface on which the initial vorticity was deposited.}
	\label{sim_M19sfi_f_cs_sim}
\end{figure}

\begin{figure}
	\centering
	\includegraphics[width=0.6\linewidth]{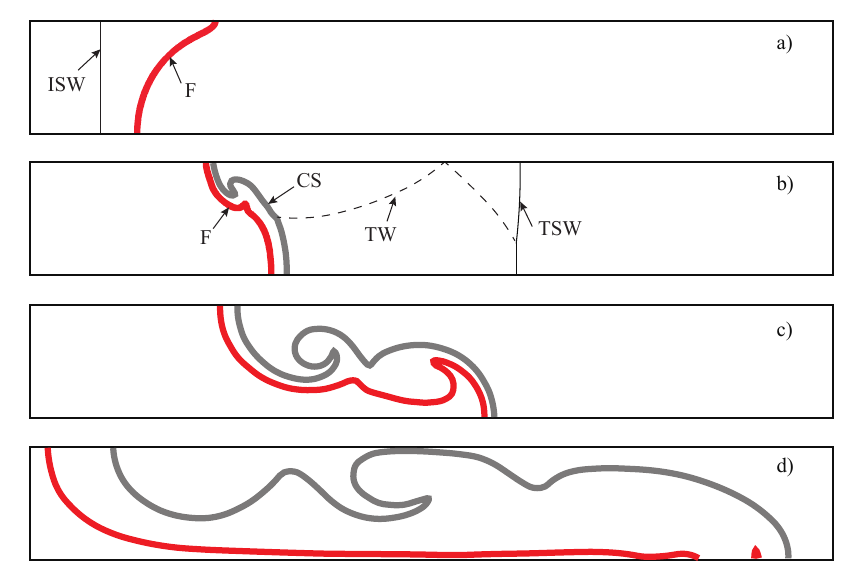}
	\caption{Sketch of the time evolution of the flame and contact surface subsequent to the passage of the incident shock illustrating the increase distance between them.}
	\label{sim_M19sfi_f_cs_sketch}
\end{figure}

A fundamental difference between inert and reactive RM instability is the location of the source of vorticity.   Both originate with the deposition of vorticity along a material surface.  In the inert case, the vorticity deposited along a material surface remains with this surface.  In the reactive case, the material surface along which vorticity has been deposited (the contact surface) separates from the new flame surface linearly in time, as new burned material accumulates between the two surfaces.  As the new flame burns away from this vorticity sheet, it consumes whatever folds have been initially formed and only largest scales persist.  This is shown in Fig.\ \ref{sim_M19sfi_f_cs_sim} and the sketch of Fig. \ref{sim_M19sfi_f_cs_sketch}. As the contact surface and its vortical flow seperates from the flame, small scale flame folds, inherited from the initial interaction, are destroyed.  As such, only the largest modes remain.  This may explain why the single mode model captures the dynamics so well. 


\subsubsection{Flame funnel collapse stage}
The non-linear reactively enhanced growth of the flame terminates when the funnels created burn out by the transverse reaction fronts. This occurs on a time scale associated with the flame's displacement velocity and the transverse length scale of the cellular flame, $\frac{\rho_b H}{\rho_u S_u}$. In the simulation, the post-interaction ratio of $\frac{\rho_u}{\rho_b}$ is found to be within the range of 6-8. With the non-dimensionalized $H$=16.5$S_u t_{fs}$, the burn-out time thus is 2.06 to 2.75$t_{fs}$, which is 1.39 to 1.85$t_{f}$. As can be observed in figure\ \ref{comp_sim_amp}, the subsequent flame relaxation to a new cellular structure occurs on this same time scale. 

\subsubsection{New cellular flame stage}

Subsequent to the collapse of the flame interface, it saturated to a new stage of cellular flame evolution. In this stage, the inert RM instability plays minor roles in influencing the flame deformation, as amplitude growth rate of which decays at a rate of $O(1/t)$ and the contact surface has propagated far from the flame. Thus, similar to the formation of the cellular structure, the flame evolution in this stage is only controlled by the intrinsic Landau-Darrieus instability. As shown in figures\ \ref{comp_sim_amp}, \ref{comp_sf_su} and \ref{su_stage4}, the growth rate of the amplitude progressively asymptoted to values close to zero except for the case of $M_s$ = 2.5 incident shock, which shows the gradual increase of positive growth rate at the end of this stage. This implies that the flames evolved to a new cellular flame evolution stage. 

\subsection{Evolution of the burning velocity}

For practical considerations, it is of interest to quantify the burning rate enhancement of the flame due to the interaction with the shock waves, and elucidate the controlling parameters. To a good approximation, the increase of the burning rate of the flame is given by the increase in its surface area \citep{sivashinsky1983instabilities}, as shown in figure \ref{steady_su_amp} for example.  For $\eta \gg H$, the flame surface area in 2D is proportional to $\eta$.  We can thus expect the characteristic rate of burning rate increase, i.e.,
\begin{equation}
R_{S_u} \equiv |\frac{d \ln S_u}{d(t/t_{f})}|
\label{equ_su_etaa}
\end{equation} 
to be well approximated by the characteristic rate of flame shape amplitude increase, i.e.,
\begin{equation}
R_{\eta} \equiv |\frac{d \ln \eta}{d(t/t_{f})}|
\label{equ_su_etab}
\end{equation}

\begin{figure}
	\centering
	\begin{subfigure}{0.495\columnwidth}
		\centering
		\includegraphics[trim={1.0cm 0cm 1cm 0cm},clip,width=1.0
		\columnwidth]{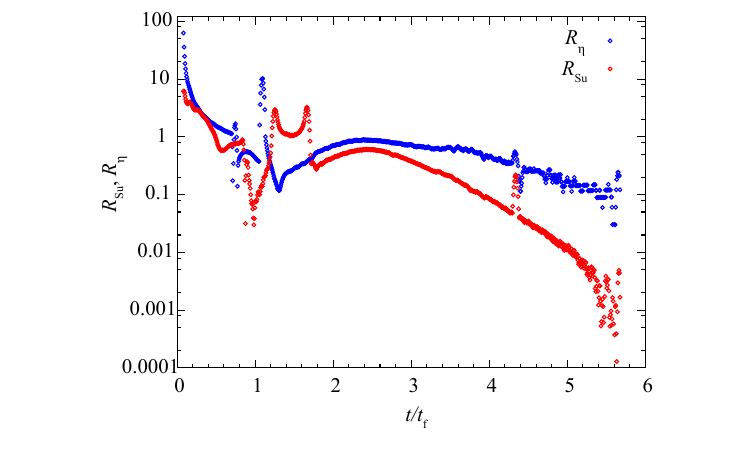}
		\subcaption{$M_s$=1.53}
		\label{}
	\end{subfigure}
	\hfill
	\begin{subfigure}{0.495\columnwidth}
		\centering
		\includegraphics[trim={1.0cm 0cm 1cm 0cm},clip,width=1.0
		\columnwidth]{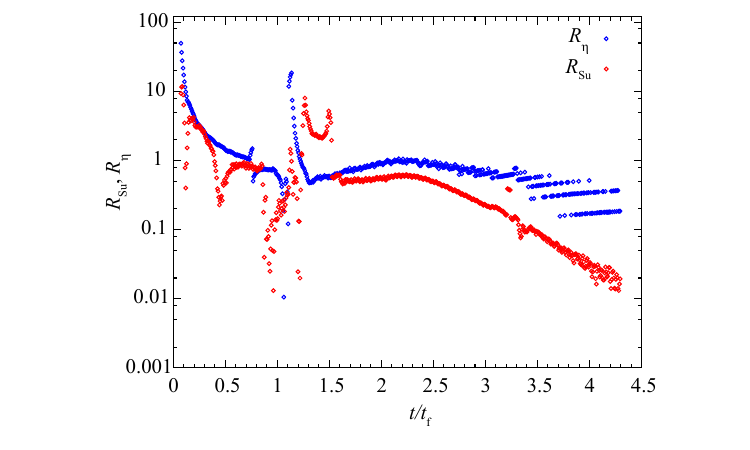}
		\subcaption{$M_s$=1.75}
		\label{}
	\end{subfigure}
	\begin{subfigure}{0.495\columnwidth}
		\centering
		\includegraphics[trim={1.0cm 0cm 1cm 0cm},clip,width=1.0
		\columnwidth]{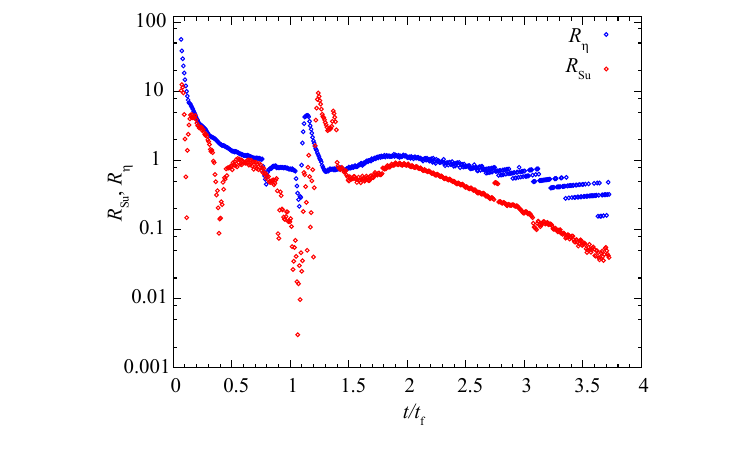}
		\subcaption{$M_s$=1.9}
		\label{}
	\end{subfigure}
	\hfill
	\begin{subfigure}{0.495\columnwidth}
		\centering
		\includegraphics[trim={1.0cm 0cm 1cm 0cm},clip,width=1.0
		\columnwidth]{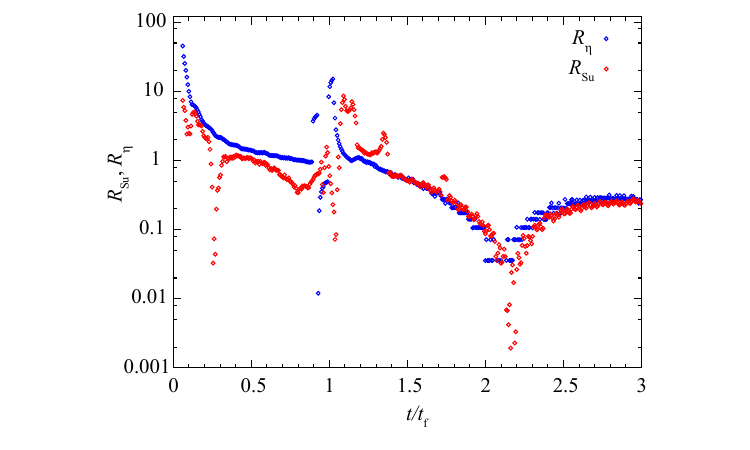}
		\subcaption{$M_s$=2.5}
		\label{}
	\end{subfigure}
	\caption{Comparison of the time evolution of burning velocity and amplitude growth rate in new cellular flame formation stage for the interaction of flame with different incident shock Mach number of 1.53, 1.75, 1.9 and 2.5.}
	\label{su_stage4}
\end{figure}

Figure\ \ref{su_stage4} shows the comparison of these two characteristic rates evaluated by taking time derivatives of the numerical signals for flame shape amplitude and burning velocity.  The two rates are found in very good agreement over the  times of interest, even for conditions where the assumption $\eta \gg H$ is not valid (at early and late times).  We have not attempted to further model the burning rate increase with higher precision given more accurate flame shape functions parametrized by the evolution of $\eta$.  This is left for future study.

\subsection{Difference between inert and reactive RM instabilities and role in DDT}
In closing, we now return to the fundamental difference between the inert and the reactive Richtmyer-Meshkov instability and the role generally attributed to the RM instability in problems of deflagration to detonation transition (DDT).   Both originate with the deposition of vorticity along a material surface.  In the inert case, this material surface subsequently deforms and eventually leads to turbulence.  In the reactive case, the material surface along which vorticity has been deposited separates from the new flame surface, as new burned material accumulates between the two surfaces.  As the influence of the vorticity deposited on this contact surface weakens as the flame surface consumes more gas, the flame burns most folds created by the initial interaction (Fig.\  \ref{sim_M19sfi_f_cs_sketch}).  The flame surface stops deforming through the action of the initially deposited vorticity, whose influence gradually decays as it separates from the flame.  As a result, the flame burns the remaining folds and returns to its original structure.  In this sense, the reactive RM instability does not generate the same level of turbulence as for the inert case. Nevertheless, the short time window available for the flame to increase its surface area, which is on the order of a few flame times, may generate a flame surface area an order of magnitude larger than prior to the interaction.  In three dimensions, the area enhancement would be larger.  Thermal-diffusive effects affecting the burning velocity of curved flames may further enhance this.  This local enhancement of flame surface area may bring the flame to the compressible regime and potentially trigger transition to detonation.

\section{Conclusion}
\label{sec:conclusion}

In this study, single shock wave colliding with a cellular flame in a shock tube was studied experimentally, numerically, and theoretically. The experiments performed in the shock tube at low pressure in stoichiometric hydrogen-air permitted to isolate and monitor the flame deformation with precision after interaction. Following the passage of the incident shock, the flame cusps were flattened and reversed backwards to the burnt gas. The flame front amplitude was taken to be an indicator for the flame surface evolution. The growth of the amplitude increased weakly with the strength of the incident Mach number. The flame propagation speed for the interaction of different Mach numbers considered with the cellular flame mildly decreased within the time of measurement. 

The simulation extended the observation time for the flame subsequent to the interaction of the experiments. Following the passage, the reversed flame interface development was found to go through the following stages: inert evolution due to the RM instability at times significantly less than the laminar flame time, nonlinear increment result from the amplification of chemical energy release on growth rate of inert RM instability, transverse burn-out of the neck of flame funnel and re-adjustment to a new cellular flame on the laminar flame time scale. A proposed model can well approximate the evolution of the flame geometry and burning rate increase for arbitrary shock strength below the shock-induced auto-ignition point and flames with Lewis number equals 1 in 2D. The model provides a simple way to estimate the flame burning rate increase upon passage of a shock and duration of this event. Future work should extend this study to incorporate wall effects and Lewis number effects, which may play pivotal role on the dynamics of the flame deformation and its coupling with the local burning rate modifications.\\

The authors acknowledge financial support provided by the Natural Sciences and Engineering Research Council of Canada (NSERC) through the Discovery Grant ”Predictability of detonation wave dynamics in gases: experiment and model development” and support of Compute Canada for the computational resources. H.Y. thanks China Scholarship Council (CSC) and NSFC grant (51774068) for the financial support. H.Y. also wants to thank Andre Fecteau for the help in calculating the 1D steady flame profile.  The authors wish to thank Sam Falle for the use of the MG code. The authors report no conflict of interest.

\appendix
\section{Measurement of the cell amplitude from experiment}
\label{sec:exp_measure}
\begin{figure}
	\centering
	\includegraphics[width=0.6\linewidth]{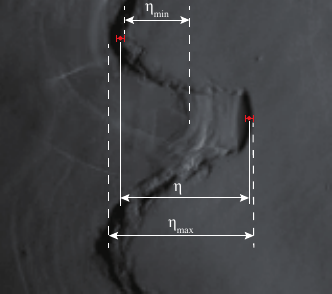}
	\caption{The density profile of the middle cell in the 6th frame of figure\ \ref{M1_9interaction} illustrating the error from the measurement.}
	\label{measure_err}
\end{figure}

Figure\ \ref{measure_err} shows the example of the measurement of the cell amplitude from experiment. The distance between the two red points, which marked by $\eta$, illustrate the leading edge of the flame front where the flame is not influenced by the wall. The red error bars represent the measurement error. The largest distance $\eta_{max}$ and the smallest distance $\eta_{min}$  account for the uncertainties due to the superposition effects of the 3D and the wall effects in the Schlieren images.

\section{Extraction of the burning velocity from experiment}
\label{appendix:xt}

A simple self-similar one-dimensional gasdynamic model was formulated in order to extract the effective burning velocity increase from the interaction of a shock with a flame.  The model uses the experimentally measured flame displacement speed to infer the corresponding burning velocity. 

\begin{figure}
	\centering
	\includegraphics[width=0.6\linewidth]{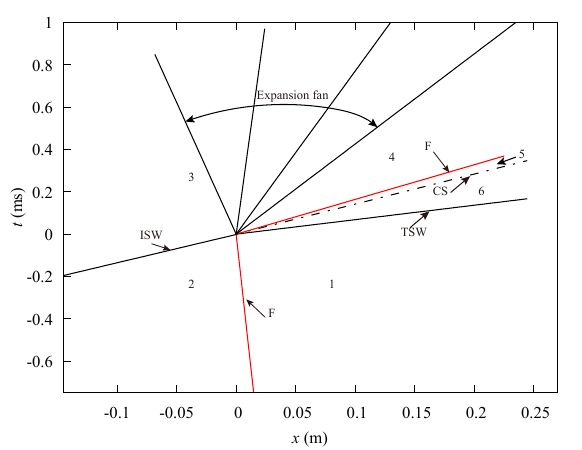}
	\caption{The space-time diagram evolution of interaction of figure\ \ref{M1_9interaction} reconstructed from the one-dimensional gas dynamic model.}
	\label{xt_reconstruction}
\end{figure}

The model assumes gasdynamic discontinuities for the incident and transmitted shocks, the flame brush and a contact surface, as illustrated in figure \ref{xt_reconstruction}.  The incident shock always transmits as a shock.  If the flame surface were inert, since the acoustic impedance $\rho c$ of the burned gases is lower than that of the unburned fresh gases owing to the large density changes, the incident shock reflects as an expansion wave.  For the reactive case, a flame surface is introduced, such that the layer of gas 5 is the gas consumed by the flame since the interaction.  All waves are assumed to have a constant speed and all regions separating the discontinuities are assumed uniform.  As such, the model is effectively a time-averaged approximation.  

Our flow reconstruction uses the knowledge of the initial states, the initial flame propagation speed and shape, the incident shock propagation speed and the displacement velocity of the flame to infer all states of interest and the flame burning velocity.  State 2 across the flame was computed from the adiabatic flame model using Cantera and the thermo-chemical database of \citet{li2004updated}. The cellular flame burning velocity was corrected from the free flame speed from Cantera by considering the ratio of the flame area and channel cross-sectional area, whereas the pressure increase in state 2 was evaluated by the method of characteristics. The flame propagation velocity before the interaction was measured from the mean average moving distance per unit time after the flame formed 3 cells. State 2 is then calculated by applying the mass and momentum conservation balance. The flow from state 2 to state 3 and state 1 to state 6 across the shock obey the usual shock jump equations. The expansion from state 4 to state 3 is assumed to be isentropic. For all states, a two-gamma approximation was used, where the non-reacted and reacted gas are assumed to be polytropic gases with constant thermodynamics properties \citep{chue1993chapman}.  A system of equations is then built from state 3 to state 6, in which the average post-shock flame propagation speed was matched with the experimentally measured value in order to infer the other flow parameters such as the flame burning velocity. 

Over the time interval available in the experiments, the burning velocity was evaluated to be 7.1$\pm$4.5m/s for the case of the $M_s$=1.9 shock-flame interaction. Note that a minor difference of propagation speed can largely influence the 1D evaluation of the burning velocity. The mean evaluated burning velocity is 3.58 times the laminar flame speed and 1.3 times the one from the simulation, which shows relatively good agreement with the simulation result. 

\section{Numerical resolution study}
\label{sec:resolution}
\begin{figure}
	\centering
	\begin{subfigure}[ht]{0.9\linewidth}
		\centering
		\includegraphics[width=1.0\linewidth]{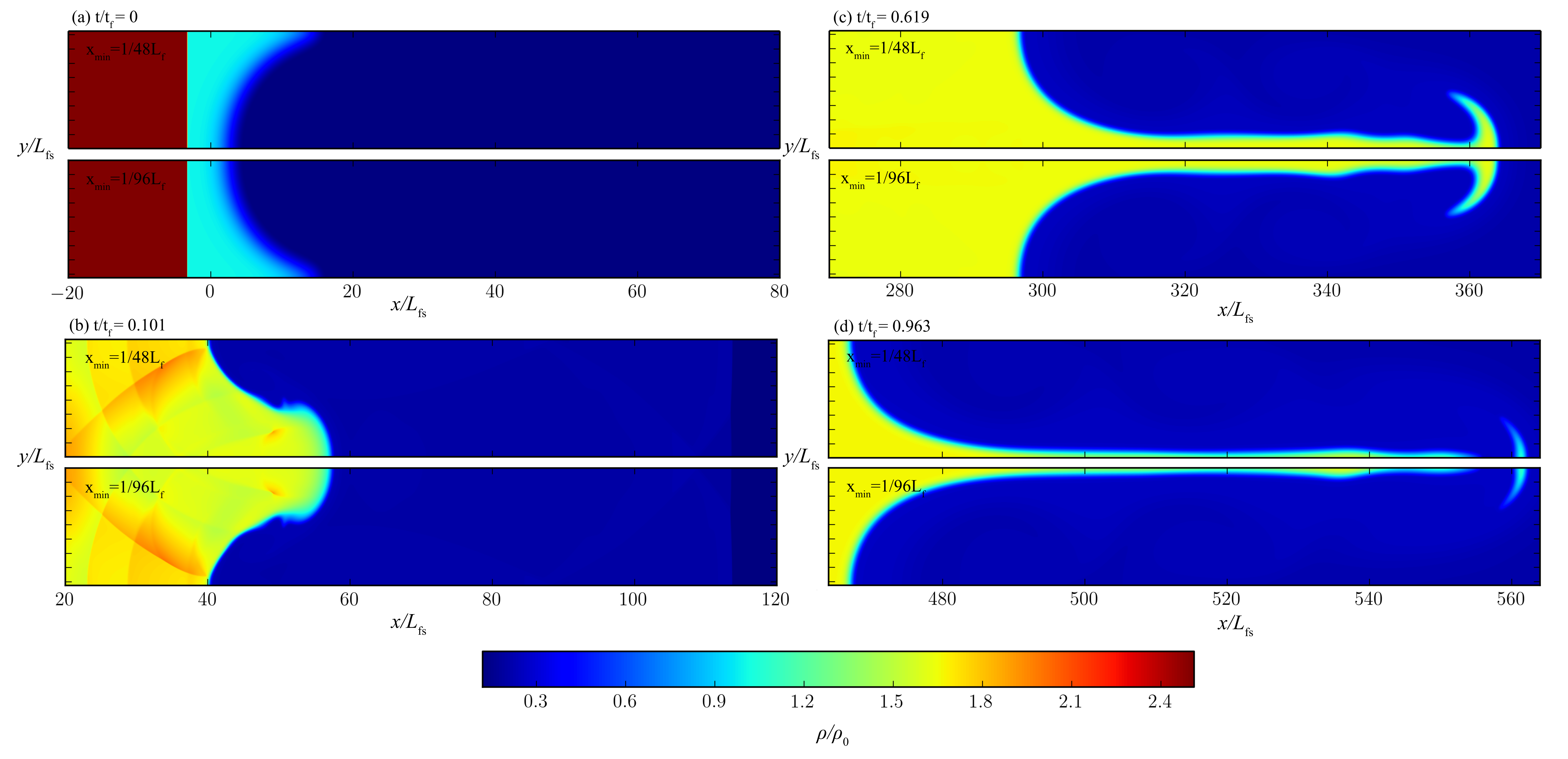}
		\subcaption{}
		\label{density-resolution_comp}
	\end{subfigure}
	\begin{subfigure}[ht]{0.7\linewidth}
		\centering
		\includegraphics[width=1.0\linewidth]{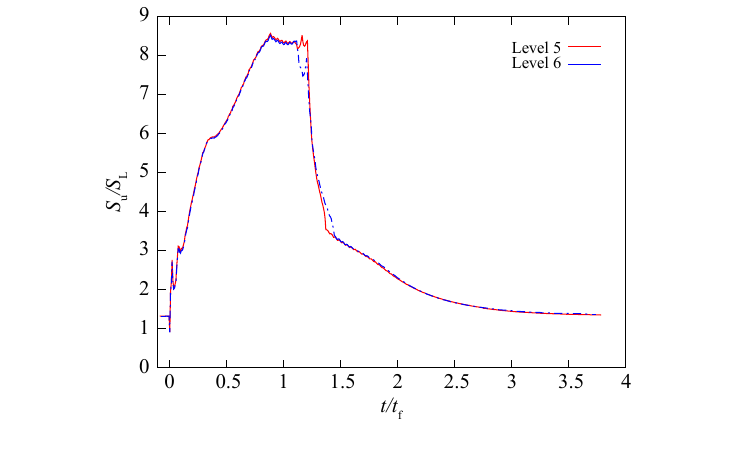}
		\subcaption{}
		\label{su_resolution_comp}
	\end{subfigure}
	\caption{Comparison of (a) density profile and (b) time evolution of burning velocity for different resolutions considered.} 
	\label{convergence_study}
\end{figure}

In order to test the effectiveness of the numerical resolution on the results presented earlier, we have performed a resolution study by varying the minimum grid spacing. As suggested in \citet{sharpe2006nonlinear}, roughly 30 grid points in the reaction zone of the steady, planar flame is the requirement to properly resolve the cellular flame. To make sure our grid points of 48 points per flame thickness (Level 5) is enough to resolve the cellular flame evolution after the interaction with the incident shock, we have performed a simulation with the minimum grid spacing of 96 points per flame thickness (Level 6) for convergence comparison. As shown in figure\ \ref{density-resolution_comp}, the flame front is approximately at the same location, and displays the same global characteristics of the front. A more quantitative appraisal of numerical convergence was conducted on the flame burning velocity. As illustrated in figure\ \ref{su_resolution_comp}, excellent convergence can be observed for the two simulations conducted with 48 and 96 points per flame thickness, except for the drastic decay in the flame funnel collapse stage. As the amplitude of the flame simulated with 48 grid points per flame thickness evolved also in good agreement with the experiment, we thus conclude that the grid resolution we employed is sufficient to resolve the flame.  

\bibliographystyle{jfm}
\bibliography{jfm-references}

\end{document}